\documentclass[12pt]{article}
\usepackage{amsmath,amssymb}
\usepackage{graphicx,psfrag,epsf}
\usepackage{enumerate}
\usepackage{natbib}
\usepackage{url} 
\usepackage{mathtools}

\usepackage{color}

\newcommand{\blind}{0}

\DeclareFontFamily{OT1}{pzc}{}
\DeclareFontShape{OT1}{pzc}{m}{it}{<-> s * [1.10] pzcmi7t}{}
\DeclareMathAlphabet{\mathpzc}{OT1}{pzc}{m}{it}
\DeclareMathOperator*{\plim}{plim}

\newcommand{\bbeta}{{\mbox{\boldmath $\beta$}}}

\newcommand{\bepsilon}{{\mbox{\boldmath $\epsilon$}}}
\newcommand{\bvarepsilon}{{\mbox{\boldmath $\varepsilon$}}}
\newcommand{\bGamma}{{\mbox{\boldmath $\Gamma$}}}

\newcommand{\y}{{\bf{{y}}}}
\newcommand{\x}{{\bf{x}}}
\newcommand{\X}{{\bf{X}}}
\newcommand{\W}{{\bf{W}}}
\newcommand{\I}{{\bf{I}}}

\newcommand{\V}{{\bf{{V}}}}
\newcommand{\bH}{{\bf{{H}}}}
\newcommand{\Z}{{\bf{{Z}}}}
\newcommand{\B}{{\bf{{B}}}}
\newcommand{\bb}{{\bf{{b}}}}
\newcommand{\ba}{{\bf{{a}}}}

\newcommand{\A}{{\bf{{A}}}}

\newcommand{\R}{{\bf{{R}}}}
\newcommand{\Q}{{\bf{{Q}}}}

\newcommand{\C}{{\bf{{C}}}}
\newcommand{\U}{{\bf{{U}}}}
\newcommand{\D}{{\bf{{D}}}}
\newcommand{\bS}{{\bf{{S}}}}
\newcommand{\bfP}{{\bf{{P}}}}

\newcommand{\bv}{{\bf{v}}}
\newcommand{\e}{{\bf{e}}}
\newcommand{\z}{{\bf{z}}}
\newcommand{\bfr}{{\bf{{r}}}}
\newcommand{\bLambda}{{\mbox{\boldmath $\Lambda$}}}

\newcommand{\bmu}{{\mbox{\boldmath $\mu$}}}
\newcommand{\bfeta}{{\mbox{\boldmath $\eta$}}}

\newcommand{\bzero}{{\mbox{\boldmath $0$}}}

\newcommand{\mcy}{{ \mbox{\boldmath $\mathpzc{y}$}}}
\newcommand{\mcX}{{\mbox{\boldmath $\mathcal{X}$}}}
\newcommand{\1}{{\bf{{1}}}}
\newtheorem{thm}{Theorem}[section]

\newtheorem{lem}[thm]{Lemma}
\newtheorem{prop}[thm]{Proposition}

\newtheorem{define}[thm]{Definition}
\newtheorem{remark}[thm]{Remark}
\def\Real{\mathbb{ R}}

\newcommand{\singlespace}{\renewcommand{\baselinestretch}{1}\small\normalsize}

\newcommand{\doublespace}{\renewcommand{\baselinestretch}{1.75}\small\normalsize}

\begin{document}

\def\spacingset#1{\renewcommand{\baselinestretch}%
{#1}\small\normalsize} \spacingset{1}


\if0\blind
{
  \title{\bf Online Updating of Statistical Inference in the Big Data Setting}
  \author{Elizabeth D. Schifano,\thanks{
    Part of the computation was done on the Beowulf cluster of the Department of Statistics, University of Connecticut, partially financed by the NSF SCREMS (Scientific Computing Research Environments for the Mathematical Sciences) grant number 0723557.
}\hspace{.2cm}\\
    Department of Statistics, University of Connecticut\\
    Jing Wu, \\
    Department of Statistics, University of Connecticut\\
    Chun Wang, \\
    Department of Statistics, University of Connecticut\\
    Jun Yan, \\
    Department of Statistics, University of Connecticut\\
    and \\
    Ming-Hui Chen \\
    Department of Statistics, University of Connecticut}
  \maketitle
} \fi

\if1\blind
{
  \bigskip
  \bigskip
  \bigskip
  \begin{center}
    {\LARGE\bf Online Updating of Statistical Inference in the Big Data Setting}
\end{center}
  \medskip
} \fi

\bigskip
\begin{abstract}
We present statistical methods for big data arising from online analytical processing, where large amounts of data arrive in streams and require fast analysis without storage/access to the historical data. In particular, we develop iterative estimating algorithms and statistical inferences for linear models and estimating equations that update as new data arrive. These algorithms are computationally efficient, minimally storage-intensive, and allow for possible rank deficiencies in the subset design matrices due to rare-event covariates. Within the linear model setting, the proposed online-updating framework leads to predictive residual tests that can be used to assess the goodness-of-fit of the hypothesized model. We also propose a new online-updating estimator under the estimating equation setting.  Theoretical properties of the goodness-of-fit tests and proposed estimators are examined in detail. In simulation studies and real data applications, our estimator compares favorably with competing approaches under the estimating equation setting.
\end{abstract}

\noindent%
\vfill

\newpage
\spacingset{1.45} 
\section{Introduction}\label{sec:intro}

The advancement and prevalence of computer technology in nearly every realm of science and daily life has enabled the collection of ``big data''.
While access to such wealth of information opens the door towards new discoveries, it also poses challenges to the current statistical and computational theory and methodology, as well as challenges for data storage and computational efficiency.

Recent methodological developments in statistics that address the big data
challenges have largely focused on subsampling-based \citep[e.g.,][]{Klei:Talw:Sark:Jord:scal:2014,
Lian:etal:resa:2013, Ma:Maho:Yu:stat:2013} and divide and conquer \citep[e.g.,][]{Lin:Xi:aggr:2011,RHIPE, ChenXie:2014} techniques; see \cite{softrev} for a review.  
``Divide and conquer'' (or ``divide and recombine'' or `split and conquer'', etc.), in particular, has become a popular approach for the analysis of large complex data. The approach is appealing because the data are first divided into subsets and then numeric and visualization methods are applied to each of the subsets separately.  The divide and conquer approach culminates by aggregating the results from each subset to produce a final solution. To date, most of the focus in the final aggregation step is in estimating the unknown quantity of interest, with little to no attention devoted to standard error estimation and inference.


In some applications,  data arrives in streams or in large
chunks, and an online, sequentially updated analysis is desirable
without storage requirements.
As far as we are aware, we are the first to examine inference in the online-updating setting.  Even with big data, inference remains an important issue for statisticians, particularly in the presence of rare-event covariates. In this work, we provide standard error formulae for divide-and-conquer estimators in the linear model (LM) and estimating equation (EE) framework.  We further develop iterative estimating algorithms and statistical inferences for the LM and EE frameworks for online-updating, which update as new data arrive.  These algorithms are computationally  efficient, minimally storage-intensive,  and allow for possible rank deficiencies in the subset design matrices due to rare-event covariates.  
  Within the online-updating setting for linear models, 
we propose tests for outlier detection based on predictive residuals and 
derive the exact distribution and the asymptotic distribution of the test statistics for the normal and non-normal cases, respectively.
 In addition, within the online-updating setting for estimating equations, we propose a new estimator and show that it is
asymptotically consistent. We further establish new uniqueness results for the resulting cumulative EE estimators in the presence of rank-deficient subset design matrices. Our simulation study and real data analysis demonstrate that the proposed estimator 
outperforms other divide-and-conquer or online-updated estimators in terms of bias and mean squared error.

The manuscript is organized as follows.  In Section \ref{sec:lm}, we first briefly review the divide-and-conquer approach for linear regression models and introduce formulae to compute the mean square error.  We then present the linear model online-updating algorithm, address possible rank deficiencies within subsets, and propose predictive residual diagnostic tests. In Section \ref{sec:ee}, we review the divide-and-conquer approach of \cite{Lin:Xi:aggr:2011} for estimating equations and introduce corresponding variance formulae for the estimators.  We then build upon this divide-and-conquer strategy to derive our online-updating algorithm and new online-updated estimator. We further provide theoretical results for the new online-updated estimator and address possible rank deficiencies within subsets.  Section \ref{sec:sim} contains our numerical simulation results for both the LM and EE settings, while Section \ref{sec:data} contains results from the analysis of real data regarding airline on-time statistics.  We conclude with a brief discussion.

\section{Normal Linear Regression Model}\label{sec:lm}

\subsection{Notation and Preliminaries}\label{sec:divandconq}
Suppose there are $N$ independent observations $\{ (y_i,\x_i), \; i=1,2,\dots,N\}$ of interest and we wish to
 fit a normal linear regression model
$y_i= \x_i'\bbeta + \epsilon_i,$  where  $\epsilon_i \sim N(0,\sigma^2)$
independently for $i=1,2,\dots,N$, and $\bbeta$ is a $p$-dimensional vector of regression coefficients corresponding to covariates $\x_i$ ($p \times 1$). Write $\y=(y_1,y_2,\dots,y_N)'$ and $\X=(\x_1,\x_2,\dots,\x_N)'$ where we assume the design matrix $\X$ is of full rank $p<N.$
The least squares (LS) estimate of $\bbeta$ and the corresponding residual mean square, or mean squared error (MSE), are given by $\hat{\bbeta} = (\X'\X)^{-1} \X'\y$ and $\mbox{MSE}=\frac{1}{N-p} \y' (\I_N -\bH)\y$, respectively,
where $\I_N$ is the $N\times N$ identity matrix and  $\bH=\X (\X'\X)^{-1}\X'.$ 

In the online-updating setting, we suppose that the $N$ observations are not available all at once, but rather arrive in chunks from a large data stream.  Suppose at each accumulation point $k$ we observe $\y_k$ and $\X_k$, the $n_k$-dimensional vector of responses and
the $n_k\times p$ matrix of covariates, respectively, for $k=1,\ldots,K$ such that $\y=(\y_1',\y_2',\dots,\y'_K)'$
and $\X=(\X_1',\X_2',\dots,\X_K')'$.
%
%
Provided $\X_k$ is of full rank, the LS estimate of $\bbeta$ based on the $k^{th}$ subset is given by
\begin{equation}
\hat{\bbeta}_{n_k, k}=(\X_k'\X_k)^{-1} \X_k'\y_k \label{betahatk}
\end{equation}
and the MSE is given by
\begin{equation}
\label{msenkk}
\mbox{MSE}_{n_k,k}=\frac{1}{n_k-p} \y'_k(\I_{n_k}- \bH_k)\y_k,
\end{equation}
where $\bH_k=\X_k(\X_k'\X_k)^{-1}\X_k'$,
for $k=1,2,\dots,K$.

As in the divide-and-conquer approach \citep[e.g.,][]{Lin:Xi:aggr:2011}, we can write $\hat{\bbeta}$ as
\begin{equation}
\hat{\bbeta}= \Big( \sum^K_{k=1} \X_k'\X_k\Big)^{-1}
 \sum^K_{k=1} \X_k'\X_k \hat{\bbeta}_{n_k, k}. \label{Ksubsetbeta}
\end{equation} We provide a similar divide-and-conquer expression for the residual sum of squares, or sum of squared errors (SSE), given by
\begin{eqnarray}
\mbox{SSE} 
& = & \; \sum^K_{k=1} \y_k'\y_k- \Big( \sum^K_{k=1} \X_k'\X_k \hat{\bbeta}_{n_k, k}\Big)'
 \Big( \sum^K_{k=1} \X_k'\X_k\Big)^{-1}
 \Big(\sum^K_{k=1} \X_k'\X_k \hat{\bbeta}_{n_k, k} \Big), \label{MSEform}
\end{eqnarray}
and $\mbox{MSE} = \mbox{SSE}/(N-p).$ The SSE, written as in \eqref{MSEform}, is quite useful if one is interested in performing inference in the divide-and-conquer setting, as $\mbox{var}(\hat\bbeta)$ may be estimated by $\mbox{MSE}(\X'\X)^{-1}=\mbox{MSE}\Big( \sum^K_{k=1} \X_k'\X_k\Big)^{-1}.$  We will see in Section \ref{sec:lmupdate} that both $\hat\bbeta$ in \eqref{Ksubsetbeta} and SSE in \eqref{MSEform} may be expressed in sequential form that is more advantageous from the perspective of online-updating.


\subsection{Online Updating}\label{sec:lmupdate}
While equations \eqref{Ksubsetbeta} and \eqref{MSEform} are quite amenable to parallel processing for each subset, the online-updating approach for data streams is inherently sequential in nature. Equations \eqref{Ksubsetbeta} and \eqref{MSEform} can certainly be used for  estimation and inference for regression coefficients resulting at some terminal point $K$ from a data stream, provided quantities $(\X_k'\X_k, \hat{\bbeta}_{n_k,k},\y_k'\y_k)$ are available for all accumulation points $k=1,\ldots,K.$  However, such data storage may not always be possible or desirable.  Furthermore, it may also be of interest to perform inference at a given accumulation step $k$, using the $k$ subsets of data observed to that point.  Thus, our objective is to formulate a computationally efficient and minimally storage-intensive procedure that will allow for online-updating of estimation and inference.

\subsubsection{Online Updating of LS Estimates}\label{sec:lsest}
While our ultimate estimation and inferential procedures are frequentist in nature, a Bayesian perspective provides some insight into how we may construct our online-updating estimators. Under a Bayesian framework, using the previous $k-1$ subsets of data to construct a prior distribution for the current data in subset $k$, we immediate identify the appropriate online updating formulae for estimating the regression coefficients $\bbeta$ and the error variance $\sigma^2$ with each new incoming dataset $(\y_k,\X_k)$.
 The Bayesian paradigm and accompanying formulae are provided in the Supplementary Material.

Let $\hat{\bbeta}_k$ and MSE$_k$ denote the LS estimate of $\bbeta$
and the corresponding MSE based on
the cumulative data $D_k=\{(\y_\ell, \X_\ell),\; \ell=1,2,\dots,k\}$.
The online-updated estimator of $\bbeta$ based on cumulative data $D_k$ is given by
\begin{equation}
\hat{\bbeta}_k=(\X_k'\X_k+\V_{k-1})^{-1} (\X_k'\X_k \hat{\bbeta}_{n_k k} + \V_{k-1} \hat{\bbeta}_{k-1}),
 \label{updatedLSbetak1}
\end{equation}
where $\hat{\bbeta}_0=\bzero$, $\hat{\bbeta}_{n_k k}$ is defined by
\eqref{betahatk} or \eqref{eq:ginvbhat}, $\V_k=\sum^k_{\ell=1} \X'_\ell \X_\ell$ for  $k=1,2,\dots$, and $\V_0=\bzero_p$ is a $p\times p$ matrix of zeros.
 Although motivated through Bayesian arguments, \eqref{updatedLSbetak1} may also be found in a (non-Bayesian) recursive linear model framework \citep[e.g.,][page 313]{stengel2012optimal}.

The online-updated estimator of the SSE based on cumulative data $D_k$ is given by
\begin{eqnarray}
\mbox{SSE}_{k}
 & = & \mbox{SSE}_{k-1} + \mbox{SSE}_{n_k,k}
 + \hat{\bbeta}_{k-1}'\V_{k-1}\hat{\bbeta}_{k-1}+ \hat{\bbeta}_{n_k, k}'\X_k'\X_k\hat{\bbeta}_{n_k, k} - \hat{\bbeta}_{k}'\V_k\hat{\bbeta}_{k}  \label{MSEupdatek.variant1}
\end{eqnarray}
where SSE$_{n_k,k}$ is the residual sum of squares from the $k^{th}$ dataset, with corresponding residual mean square MSE$_{n_k,k}=$SSE$_{n_k,k}/(n_k-p).$
The MSE based on the data $D_k$ is then $ \mbox{MSE}_k=\mbox{SSE}_k/(N_k-p)$
where $N_k=\sum^k_{\ell=1} n_\ell$ $(=n_k+N_{k-1})$ for $k=1,2,\ldots$.  Note that for $k=K$, equations \eqref{updatedLSbetak1} and \eqref{MSEupdatek.variant1} are identical to those in \eqref{Ksubsetbeta} and \eqref{MSEform}, respectively.

Notice that, in addition to quantities only involving the current data $(\y_k,\X_k)$ (i.e., $\hat{\bbeta}_{n_k, k}$, $\mbox{SSE}_{n_k,k}$, $\X_k'\X_k,$ and $n_k$),  we only used quantities $(\hat{\bbeta}_{k-1},\mbox{SSE}_{k-1}, \V_{k-1}, N_{k-1})$ from the previous accumulation point to compute $\hat{\bbeta}_{k}$ and $\mbox{MSE}_k.$  Based on these online-updated estimates, one can easily obtain online-updated t-tests for the regression parameter estimates. Online-updated ANOVA tables require storage of two additional scalar quantities from the previous accumulation point; details are provided in the Supplementary Material.

\subsubsection{Rank Deficiencies in $\X_k$}\label{sec:lm.rank}
When dealing with subsets of data, either in the divide-and-conquer or the online-updating setting, it is quite possible (e.g., in the presence of rare event covariates) that some of the design matrix subsets $\X_k$ will not be of full rank, even if the design matrix $\X$ for the entire dataset is of full rank.  For a given subset $k$, note that if the columns of $\X_k$ are not linearly independent, but lie in a space of dimension $q_k<p$, the estimate
\begin{equation}\label{eq:ginvbhat}
\hat{\bbeta}_{n_k, k} = (\X_k'\X_k)^{-} \X_k'\y_k,
\end{equation}
where $(\X_k'\X_k)^{-}$ is a generalized inverse of $(\X_k'\X_k)$ for subset $k$, will not be unique.  However, both $\hat\bbeta$ and MSE will be unique, which leads us to introduce the following proposition.

\begin{prop}\label{prop:lm}
Suppose $\X$ is of full rank $p<N$.  If the columns of $\X_k$ are not linearly independent, but lie in a space of dimension $q_k<p$ for any $k=1,\ldots,K$, $\hat\bbeta$ in \eqref{Ksubsetbeta} and SSE \eqref{MSEform} using $\hat{\bbeta}_{n_k, k}$ as in \eqref{eq:ginvbhat} will  be invariant to the choice of generalized inverse $(\X_k'\X_k)^{-}$ of $(\X_k'\X_k)$.
\end{prop}

To see this, recall that a generalized inverse of a matrix $\B$, denoted by $\B^{-},$ is a matrix such that $\B\B^{-}\B=\B.$  Note that for $(\X_k'\X_k)^{-},$ a generalized inverse of $(\X_k'\X_k),$  $\hat{\bbeta}_{n_k, k}$ given in \eqref{eq:ginvbhat} is a solution to the linear system $(\X_k'\X_k)\bbeta_k=\X_k'\y_k.$  It is well known that if $(\X_k'\X_k)^{-}$ is a generalized inverse of $(\X_k'\X_k),$ then $\X_k(\X_k'\X_k)^{-}\X_k'$ is invariant to the choice of $(\X_k'\X_k)^{-}$ \citep[e.g.,][p20]{searle:1971}. Both \eqref{Ksubsetbeta} and \eqref{MSEform} rely on $\hat{\bbeta}_{n_k, k}$ only through product
$\X_k'\X_k \hat{\bbeta}_{n_k, k}= \X_k'\X_k (\X_k'\X_k)^{-} \X_k'\y_k = \X_k'\y_k $
which is invariant to the choice of $(\X_k'\X_k)^{-}.$  

\begin{remark}\label{rem:invert}
The online-updating formulae \eqref{updatedLSbetak1} and \eqref{MSEupdatek.variant1} do not require $\X_k'\X_k$ for all $k$ to be invertible.  In particular, the online-updating scheme only requires $\V_k=\sum_{\ell=1}^k \X_\ell'\X_\ell$ to be invertible.  This fact can be made more explicit by rewriting \eqref{updatedLSbetak1} and \eqref{MSEupdatek.variant1}, respectively, as
\begin{eqnarray} 
  \label{eq:noinvbeta}
 \hat{\bbeta}_k &=& (\X_k'\X_k+\V_{k-1})^{-1} (\X_k'\y_k + \W_{k-1}) = \V_{k}^{-1} (\X_k'\y_k + \W_{k-1})\\
  \label{eq:noinv}
 \mbox{SSE}_k &=& \mbox{SSE}_{k-1} + \y_k'\y_k + \hat\bbeta_{k-1}'\V_{k-1}\hat\bbeta_{k-1} - \hat\bbeta_k'\V_k\hat\bbeta_k
 \end{eqnarray}
where $\W_0=\bzero$ and $\W_{k}=\sum_{\ell=1}^k \X_{\ell}'\y_{\ell}$ for $k=1,2,\ldots$.  
\end{remark}

\begin{remark}\label{rem:ridge}
Following Remark \ref{rem:invert} and using the Bayesian motivation discussed in the Supplementary Material, if $\X_1$ is not of full rank (e.g., due to a rare event covariate), we may consider a regularized least squares estimator by setting $\V_0\neq \bzero_p.$  For example, setting $\V_0=\lambda\I_p, \lambda>0,$ with $\bmu_0=\bzero$ would correspond to a ridge estimator and could be used at the beginning of the online estimation process until enough data has accumulated; once enough data has accumulated, the biasing term $\V_0=\lambda\I_p$ may be removed such that the remaining sequence of updated estimators $\hat\bbeta_k$ and MSE$_k$ are unbiased for $\bbeta$ and $\sigma^2,$ respectively.  More specifically, set $\V_k = \sum_{\ell=0}^k \X_\ell'\X_\ell$ (note that the summation starts at $\ell=0$ rather than $\ell=1$) where $\X_0'\X_0\equiv \V_0$, keep $\hat\bbeta_0 = \bzero,$ and suppose at accumulation point $\kappa$ we have accumulated enough data such that $\X_{\kappa}$ is of full rank.  For $k<\kappa$ and $\V_0=\lambda\I_p, \lambda>0,$ we obtain a (biased) ridge estimator and corresponding sum of squared errors by using \eqref{updatedLSbetak1} and \eqref{MSEupdatek.variant1} or \eqref{eq:noinvbeta} and \eqref{eq:noinv}.  At $k=\kappa$, we can remove the bias with, e.g.,
\begin{eqnarray} 
 \hat{\bbeta}_{\kappa} &=& (\X_\kappa'\X_\kappa+\V_{\kappa-1}-\V_0)^{-1} (\X_\kappa'\y_\kappa + \W_{\kappa-1}) \\
 \mbox{SSE}_{\kappa} &=& \mbox{SSE}_{\kappa-1} + \y_\kappa'\y_\kappa + \hat\bbeta_{\kappa-1}'\V_{\kappa-1}\hat\bbeta_{\kappa-1} - \hat\bbeta_\kappa'(\V_\kappa-\V_0)\hat\bbeta_\kappa ,
 \end{eqnarray}
and then proceed with original updating procedure for $k>\kappa$ to obtain unbiased estimators of $\bbeta$ and $\sigma^2.$
\end{remark}

\subsection{Model Fit Diagnostics}\label{sec:modfit.diag}
While the advantages of saving only lower-dimensional summaries are clear, a potential disadvantage arises in terms of difficulty performing classical residual-based model diagnostics.  Since we have not saved the individual observations from the previous $(k-1)$ datasets, we can only compute residuals based upon the current observations $(\y_k, \X_k).$  For example, one may compute the residuals
$e_{ki} = y_{ki}-\hat{y}_{ki},$ where $i=1,\ldots, n_k$ and
 $\hat{y}_{ki}=\x_{ki}'\hat{\bbeta}_{n_k,k}$, or even the externally studentized residuals given by
\begin{equation}\label{eq.e.stud}
t_{ki}=\frac{e_{ki}}{ \sqrt{ {\rm MSE}_{n_k,k(i)} (1-h_{k,ii})}}=e_{ki} \Big[ \frac{n_k-p-1}{ {\rm SSE}_{n_k,k} (1-h_{k,ii})-e^2_{ki}} \Big]^{1/2},
\end{equation}
where $h_{k,ii}=\mbox{Diag}(\bH_k)_i=\mbox{Diag}(\X_k(\X_k'\X_k)\X_k')_i$ and MSE$_{n_k,k(i)}$ is the MSE computed from the $k^{th}$ subset with the $i^{th}$ observation removed, $i=1,\ldots, n_k.$

However, for model fit diagnostics in the online-update setting, it would arguably be more useful to consider the \textit{predictive residuals}, based on $\hat{\bbeta}_{k-1}$ from data $D_{k-1}$ with predicted values $\check{\y}_{k}=(\check{y}_{k1},\ldots,\check{y}_{kn_k})'=\X_k\hat{\bbeta}_{k-1},$ as
$
\check{e}_{ki} = y_{ki} - \check{y}_{ki},~~~i=1,\ldots, n_k.
$
Define the standardized predictive residuals as
\begin{equation}\label{eq:indiv.res}
\check{t}_{ki} = \check{e}_{ki} / \sqrt{\hat{\mbox{var}}(\check{e}_{ki})},~~~ i=1,\ldots,n_k.
\end{equation}

\subsubsection{Distribution of standardized predictive residuals}\label{sec:dist.pr}
To derive the distribution of $\check{t}_{ki},$ we introduce new notation. Denote $\mcy_{k-1} = (\y_1',\ldots,\y_{k-1}')',$ and $\mcX_{k-1}$ and $\bvarepsilon_{k-1}$ the corresponding $N_{k-1}\times p$ design matrix of stacked $\X_\ell,$ $\ell=1,\ldots, k-1,$ and $N_{k-1}\times 1$ random errors, respectively. For new observations $\y_k,\X_k$, we assume
\begin{equation}\label{eq:klinmod}
 \y_k = \X_k\bbeta + \bepsilon_k ,
 \end{equation}
 where the elements of $\bepsilon_{k}$ are independent with mean 0 and variance $\sigma^2$ independently of the elements of $\bvarepsilon_{k-1}$ which also have mean 0 and variance $\sigma^2.$
Thus,
$E(\check{e}_{ki}) = 0,$ $\mbox{var}(\check{e}_{ki}) =  \sigma^2(1+x_{ki}'( \mcX_{k-1}'\mcX_{k-1})^{-1} x_{ki})$
for $i=1,\ldots,n_k$, and
$$\mbox{var}(\check{\e}_k) =\sigma^2(\I_{n_k} + \X_k( \mcX_{k-1}'\mcX_{k-1})^{-1}\X_k')$$
 where $\check{\e}_k=(\check{e}_{k1},\ldots,\check{e}_{kn_k})'.$

  If we assume that both $\bepsilon_{k}$ and $\bvarepsilon_{k-1}$ are normally distributed, then it is easy to show that $\check{\e}_{k}'\mbox{var}(\check{\e}_k)^{-1}\check{\e}_{k}\sim \chi^2_{n_k}.$ Thus, estimating $\sigma^2$ with MSE$_{k-1}$ and noting that $\frac{N_{k-1} -p}{\sigma^2}\mbox{MSE}_{k-1}\sim \chi^2_{N_{k-1}-p}$ independently of $\check{\e}_{k}'\mbox{var}(\check{\e}_k)^{-1}\check{\e}_{k}$, we find that $\check{t}_{ki} \sim t_{N_{k-1}-p}$
and
\begin{equation}\label{eq:joint.predres}
 \check{F}_k:=\frac{\check{\e}_{k}'(\I_{n_k} + \X_k( \mcX_{k-1}'\mcX_{k-1})^{-1}\X_k')^{-1}\check{\e}_{k}}{n_k\mbox{MSE}_{k-1}} \sim F_{n_k, N_{k-1}-p}.
 \end{equation}
If we are not willing to assume normality of the errors, we introduce the following proposition. The proof of the proposition is given in the Supplementary Material.  
 \begin{prop}\label{prop:chisq}
Assume that
\begin{enumerate}
\item $\epsilon_i,$ $i=1,\ldots,n_k$, are independent and identically distributed with $E(\epsilon_i)=0$ and $E(\epsilon_i^2)=\sigma^2$; 
\item the elements of the design matrix $\mcX_{k}$ are uniformly bounded, i.e., $|X_{ij}|<C$, $\forall$ $i, j$, where $C<\infty$ is constant;
\item ${\displaystyle \lim_{N_{k-1} \rightarrow \infty}}\frac{\mcX_{k-1}'\mcX_{k-1}}{N_{k-1}}=\Q$, where $\Q$ is a positive definite matrix.
\end{enumerate}
Let $\check{\e}^*_k=\bGamma^{-1}\check{\e}_k$, where $\bGamma\bGamma'\triangleq\I_{n_k}+\X_k(\mcX_{k-1}'\mcX_{k-1})^{-1}\X_{k}'$. 
Write  $\check{\e}{^*_{k}}'=(\check{\e}{^*_{k_1}}', \dots, \check{\e}{^*_{k_m}}')$, where $\check{\e}^*_{k_i}$ is an $n_{k_i} \times 1$ vector consisting of the $(\sum_{\ell=1}^{i-1} n_{k_\ell} +1)$th component through
the $(\sum_{\ell=1}^{i} n_{k_\ell})$th component of $\check{\e}{^*_{k}},$ and $\sum_{i=1}^m n_{k_i}=n_k$.  We further assume that 
\begin{itemize}
\item[4.] $ \displaystyle \lim_{n_k \rightarrow \infty}\frac{n_{k_i}}{n_k}=C_i$, where $0<C_i<\infty$ is constant for $i=1, \ldots, m$.
\end{itemize}
 Then at accumulation point $k$, we have
 \begin{equation}\label{eq:joint.predres.asympt}
\frac{\sum_{i=1}^m \frac{1}{n_{k_i}}( {\bf 1}_{k_i} '\check{\e}^*_{k_i})^2}{\mbox{MSE}_{k-1}} \xrightarrow[]{d} \chi^2_{m}, \text{ \qquad as $n_k, N_{k-1} \rightarrow \infty$,}
 \end{equation}
 where ${\bf 1}_{k_i}$ is an  $n_{k_i} \times 1$ vector of all ones.
 \end{prop}

\subsubsection{Tests for Outliers}\label{sec:outlier.test}
Under normality of the random errors, we may use statistics $\check{t}_{ki}$ in \eqref{eq:indiv.res} and $\check{F}_k$ in \eqref{eq:joint.predres} to test individually or globally if there are any
outliers in the $k^{th}$ dataset. Notice that $\check{t}_{ki}$ in \eqref{eq:indiv.res} and $\check{F}_k$ in \eqref{eq:joint.predres} can be re-expressed equivalently as
\begin{eqnarray}\label{eq:t.test}
\check{t}_{ki} &=& \check{e}_{ki}/\sqrt{\mbox{MSE}_{k-1}(1+x_{ki}'(\V_{k-1})^{-1}x_{ki})}\\
\label{eq:f.test}
\check{F}_k &=& \frac{\check{\e}_{k}'(\I_{n_k} + \X_k( \V_{k-1})^{-1}\X_k')^{-1}\check{\e}_{k}}{n_k\mbox{MSE}_{k-1}}
\end{eqnarray}
and thus can both be computed with the lower-dimensional stored summary statistics from the previous accumulation point.

We may identify as outlying $y_{ki}$ observations those cases whose standardized predicted $\check{t}_{ki}$ are large in magnitude.
If the regression model is appropriate, so that no case is outlying because of a change in the model, then each $\check{t}_{ki}$ will follow the $t$ distribution with $N_{k-1}-p$ degrees of freedom.  Let $p_{ki}=P(|t_{N_{k-1}-p}|>|\check{t}_{ki}|)$ be the unadjusted $p$-value and let $\tilde{p}_{ki}$ be the corresponding  \textit{adjusted} $p$-value for multiple testing \citep[e.g.,][]{Benjamini1995, BY2001}.  We will declare $y_{ki}$ an outlier if $\tilde{p}_{ki}<\alpha$ for a prespecified $\alpha$ level.  Note that while the Benjamini-Hochberg procedure assumes the multiple tests to be independent or positively correlated, the predictive residuals will be approximately independent as the sample size increases.  Thus, we would expect the false discovery rate to be controlled with the Benjamini-Hochberg $p$-value adjustment for large $N_{k-1}.$

To test if there is at least one outlying value based upon null hypothesis $H_0: E(\check{\e}_{k})=\bzero,$ we will use statistic $\check{F}_k.$ Values of the test statistic larger than $F(1-\alpha,n_k,N_{k-1}-p)$ would indicate at least one outlying $y_{ki}$ exists among $i=1,\ldots,n_k$ at the corresponding $\alpha$ level.

If we are unwilling to assume normality of the random errors, we may still perform a global outlier test under the assumptions of   
Proposition \ref{prop:chisq}.  Using Proposition \ref{prop:chisq} and following the calibration proposed in Muirhead (1982) \citep[][page 218]{muirhead2009aspects}, we obtain an asymptotic F statistic
 \begin{equation}\label{eq:f.test.asymp}
\check{F}_k^a:=\frac{\sum_{i=1}^m \frac{1}{n_{k_i}}({\bf 1}_{k_i} '\check{\e}^*_{k_i})^2}{\mbox{MSE}_{k-1}}\frac{N_{k-1}-m+1}{N_{k-1} \cdot m} \xrightarrow[]{d} F(m, N_{k-1}-m+1), \text{ ~~ as $n_k, N_{k-1} \rightarrow \infty$.}
\end{equation}  Values of the test statistic $\check{F}_k^a$ larger than $F(1-\alpha,m,N_{k-1}-m+1)$ would indicate at least one outlying observation exists among $\y_k$ at the corresponding $\alpha$ level.

\begin{remark} 
Recall that $\mbox{var}(\check{\e}_{k})=(\I_{n_k}+\X_k(\mcX_{k-1}'\mcX_{k-1})^{-1}\X_{k}')\sigma^2\triangleq \bGamma\bGamma'\sigma^2,$ where $\bGamma$ is an $n_k \times n_k$ invertible matrix. For large $n_k$, it may be challenging to compute the Cholesky decomposition of $\mbox{var}(\check{\e}_{k}).$ One possible solution that avoids the large $n_k$ issue is given in the Supplementary Material.
\end{remark}

\section{Online Updating for Estimating Equations}\label{sec:ee}

A nice property in the normal linear regression model setting is that regardless of whether one ``divides and conquers'' or performs online updating, the final solution $\hat{\bbeta}_K$ will be the same as it would have been if one could fit all of the data simultaneously and obtained $\hat\bbeta$ directly.  However, with generalized linear models and estimating equations, this is typically not the case, as the score or estimating functions are often nonlinear in $\bbeta.$  Consequently, divide and conquer strategies in these settings often rely on some form of linear approximation to attempt to convert the estimating equation
problem into a least square-type problem.  For example, following \cite{Lin:Xi:aggr:2011},
suppose $N$ independent observations $\{ \z_{i}, \; i=1,2,\dots,N\}.$  For generalized linear models, $\z_{i}$ will be $(y_i,\x_i)$ pairs, $i=1,\ldots, N$ with $E(y_i)=g(\x_i'\bbeta)$ for some known function $g$. Suppose there exists $\bbeta_0 \in \Real^p$ such that $\sum_{i=1}^N E[\psi(\z_i,\bbeta_0)]=0$ for some score or estimating function $\psi$.  Let $\hat{\bbeta}_N$ denote the solution to the estimating equation (EE)
$$
M(\bbeta) = \sum_{i=1}^{N} \psi(\z_i, \bbeta)=\bzero
$$
and let $\hat\V_N$ be its corresponding estimate of covariance, often of sandwich form.

Let $\{\z_{ki}, i = 1,\ldots,n_k\}$ be the observations in the $k$th subset.
The estimating function for subset $k$ is
\begin{equation}
  \label{eq:M}
  M_{n_k,k}(\bbeta) = \sum_{i=1}^{n_k} \psi(\z_{ki}, \bbeta).
\end{equation}
Denote the solution to $M_{n_k,k}(\bbeta) = \bzero$ as $\hat\bbeta_{n_k,k}$.
If we define
\begin{equation}
  \label{eq:A}
\A_{n_k,k} = - \sum_{i=1}^{n_k} \frac{\partial \psi(\z_{ki}, \hat\bbeta_{n_k,k})}{\partial\bbeta},
\end{equation} a Taylor Expansion of $-M_{n_k,k}(\bbeta)$ at $\hat\bbeta_{n_k,k}$ is given by
\begin{eqnarray*}
 -M_{n_k,k}(\bbeta) &=& \A_{n_k,k}(\bbeta-\hat\bbeta_{n_k,k}) + \R_{n_k,k}
\end{eqnarray*}
as $M_{n_k,k}(\hat\bbeta_{n_k,k})=\bzero$ and $\R_{n_k,k}$ is the remainder term.  As in the linear model case, we do not require $\A_{n_k,k}$ to be invertible for each subset $k$, but do require that $\sum_{\ell=1}^k \A_{n_\ell,\ell}$ is invertible.  Note that for the asymptotic theory in Section \ref{sec:asymptotics}, we assume that $\A_{n_k,k}$ is invertible for large $n_k$. For ease of notation, we will assume for now that each $\A_{n_k,k}$ is invertible, and we will address rank deficient $\A_{n_k,k}$ in Section \ref{sec:rank} below.

The aggregated estimating equation (AEE) estimator of \cite{Lin:Xi:aggr:2011} combines the subset estimators
 through
\begin{equation}
  \label{eq:aee}
  \hat\bbeta_{NK} = \left( \sum_{k=1}^K \A_{n_k,k} \right)^{-1}\sum_{k=1}^K \A_{n_k,k}  \hat\bbeta_{n_k,k}
\end{equation}
which is the solution to $\sum_{k=1}^K \A_{n_k,k}(\bbeta-\hat\bbeta_{n_k,k})=\bzero.$
 \cite{Lin:Xi:aggr:2011} did not discuss a variance formula, but a natural
variance estimator is given by
\begin{equation}
  \label{eq:vb}
  \hat \V_{NK} = \left( \sum_{k=1}^K \A_{n_k,k}\right)^{-1}\sum_{k=1}^K \A_{n_k,k} \hat \V_{n_k,k} \A_{n_k,k}^{\top} \left[\left( \sum_{k=1}^K \A_{n_k,k}\right)^{-1}\right]^{\top},
\end{equation}
where $\hat \V_{n_k,k}$ is the variance estimator of $\hat\bbeta_{n_k,k}$
from the subset $k$.
If $\hat \V_{n_k,k}$ is of sandwich form, it can be expressed as
$\A_{n_k,k}^{-1} \hat{\Q}_{n_k,k} \A_{n_k,k}^{-1}$, where $\hat{\Q}_{n_k,k}$ is an estimate of $\Q_{n_k,k} = \mbox{var}(M_{n_k,k}(\bbeta))$.
Then, the variance estimator becomes
\begin{equation}\label{eq:aeevar}
  \hat \V_{NK} = \left( \sum_{k=1}^K \A_{n_k,k}\right)^{-1}\sum_{k=1}^K \hat{\Q}_{n_k,k}  \left[\left( \sum_{k=1}^K \A_{n_k,k}\right)^{-1}\right]^{\top},
\end{equation}
which is still of sandwich form.

\subsection{Online Updating}
Now consider the online-updating perspective in which we would like to update the estimates of $\bbeta$ and its variance as new data arrives.  For this purpose, we introduce the cumulative estimating equation (CEE) estimator for the regression coefficient vector at accumulation point $k$ as
\begin{equation}
  \label{eq:upbeta}
  \hat\bbeta_{k} = (\A_{k-1} + \A_{n_{k},k})^{-1}(\A_{k-1} \hat\bbeta_{k-1} + \A_{n_{k},k} \hat\bbeta_{n_{k},k}).
\end{equation}
for $k=1,2,\dots$ where $\hat{\bbeta}_0=\bzero$, $\A_0=\bzero_p,$ and $\A_k=\sum^k_{\ell=1} \A_{n_\ell,\ell}=\A_{k-1} + \A_{n_{k},k}.$

For the variance estimator at the $k^{th}$ update, we take
\begin{equation}
  \label{eq:upvar}
 \hat \V_{k} = (\A_{k-1} + \A_{n_{k},k})^{-1} ( \A_{k-1} \hat \V_{k-1}  \A_{k-1}^{\top} + \A_{n_{k},k} \hat \V_{n_{k}, k} \A_{n_{k}, k}^{\top})[(\A_{k-1} + \A_{n_{k}, k})^{-1}]^{\top},
\end{equation}
with $\hat \V_0=\bzero_p$ and $\A_0=\bzero_p.$

By induction, it can be shown that~\eqref{eq:upbeta} is
equivalent to the AEE combination~\eqref{eq:aee} when $k=K$, and likewise \eqref{eq:upvar} is equivalent to \eqref{eq:aeevar} (i.e., AEE=CEE).  However, the AEE estimators, and consequently the CEE estimators, are not identical to the EE estimators $\hat\bbeta_N$ and $\hat\V_N$ based on all $N$ observations.  It should be noted, however, that \cite{Lin:Xi:aggr:2011} did prove asymptotic consistency of AEE estimator $\hat\bbeta_{NK}$ under certain regularity conditions.  Since the CEE estimators are not identical to the EE estimators in finite sample sizes, there is room for improvement.

Towards this end, consider the Taylor expansion of $-M_{n_k,k}(\bbeta)$ around some vector $\check\bbeta_{n_k,k},$ to be defined later. Then
\begin{eqnarray*}
 -M_{n_k,k}(\bbeta) &=& -M_{n_k,k}(\check\bbeta_{n_k,k}) + [\A_{n_k,k}(\check\bbeta_{n_k,k})](\bbeta-\check\bbeta_{n_k,k}) + \check\R_{n_k,k}
\end{eqnarray*}
with $\check\R_{n_k,k}$ denoting the remainder. Denote $\tilde{\bbeta}_K$ as the solution of
\begin{eqnarray}\label{eq:beta.tilde.K}
\sum_{k=1}^K -M_{n_k,k}(\check\bbeta_{n_k,k}) + \sum_{k=1}^K[\A_{n_k,k}(\check\bbeta_{n_k,k})](\bbeta-\check\bbeta_{n_k,k})=\bzero.
\end{eqnarray}
Define $\tilde\A_{n_k,k}=[\A_{n_k,k}(\check\bbeta_{n_k,k})]$ and assume $\A_{n_{k},k}$ refers to $\A_{n_{k},k}(\hat\bbeta_{n_k,k}).$
Then we have
\begin{equation}\label{eq:solution}
\tilde{\bbeta}_K = \left\{\sum_{k=1}^K\tilde\A_{n_k,k}\right\}^{-1}\left\{\sum_{k=1}^K\tilde\A_{n_k,k}\check\bbeta_{n_k,k} + \sum_{k=1}^KM_{n_k,k}(\check{\bbeta}_{n_k,k})\right\}.
\end{equation}

If we choose $\check\bbeta_{n_k,k}=\hat\bbeta_{n_k,k},$ then $\tilde\bbeta_K$ in \eqref{eq:solution} reduces to the AEE estimator of \cite{Lin:Xi:aggr:2011} in \eqref{eq:aee}, as \eqref{eq:beta.tilde.K} reduces to $\sum_{k=1}^K\A_{n_k,k}(\bbeta-\hat\bbeta_{n_k,k})=\bzero$ because $M_{n_k,k}(\hat\bbeta_{n_k,k})=\bzero$ for all $k=1,\ldots,K.$ However, one does not need to choose $\check\bbeta_{n_k,k}=\hat\bbeta_{n_k,k}.$ In the online-updating setting, at each accumulation point $k$, we have access to the summaries from the previous accumulation point $k-1,$ so we may use this information to our advantage when defining $\check\bbeta_{n_k,k}$.  Consider the intermediary estimator given by
\begin{equation}\label{eq:checkbeta}
 \check\bbeta_{n_k,k}= (\tilde{\A}_{k-1} + \A_{n_{k},k})^{-1}
(\sum_{\ell=1}^{k-1}\tilde\A_{n_\ell,\ell}\check\bbeta_{n_\ell,\ell} + \A_{n_{k},k} \hat\bbeta_{n_{k},k})
 \end{equation}
for $k=1,2,\ldots,$ $\tilde \A_0 =\bzero_p,$ $\check\bbeta_{n_0,0}=\bzero,$ and $\tilde\A_k = \sum_{\ell=1}^k\tilde\A_{n_\ell,\ell}.$  Estimator \eqref{eq:checkbeta} combines the previous intermediary estimators $\check\bbeta_{n_\ell,\ell},$ $\ell=1,\ldots, k-1$ and the current subset estimator $\hat\bbeta_{n_{k},k},$ and arises as the solution to the estimating equation
$$ \sum_{\ell=1}^{k-1}\tilde\A_{n_\ell,\ell}(\bbeta-\check\bbeta_{n_\ell,\ell}) + \A_{n_{k},k}(\bbeta-\hat\bbeta_{n_{k},k})=\bzero, $$
where $\A_{n_{k},k}(\bbeta-\hat\bbeta_{n_{k},k})$ serves as a bias correction term due to the omission of $-\sum_{\ell=1}^{k-1} M_{n_k,k}(\check\bbeta_{n_k,k})$ from the equation.

With the choice of $\check\bbeta_{n_k,k}$ as given in \eqref{eq:checkbeta}, we introduce the cumulatively updated estimating equation (CUEE) estimator $\tilde{\bbeta}_k$ as
\begin{eqnarray}\label{eq:cuee}
  \tilde\bbeta_{k} &=& (\tilde\A_{k-1} + \tilde{\A}_{n_{k},k})^{-1}(\ba_{k-1} + \tilde\A_{n_{k},k}\check\bbeta_{n_k,k}+ \bb_{k-1} + M_{n_k,k}(\check\bbeta_{n_k,k}))%
\end{eqnarray}
with
$
\ba_k = \sum_{\ell=1}^k \tilde{\A}_{n_{k},k}\check\bbeta_{n_k,k} = \tilde{\A}_{n_{k},k}\check\bbeta_{n_k,k} + \ba_{k-1}$ and
$\bb_k = \sum_{\ell=1}^k M_{n_k,k}(\check\bbeta_{n_k,k})=M_{n_k,k}(\check\bbeta_{n_k,k})+\bb_{k-1}
$
where  $\ba_0=\bb_0=\bzero$, $\tilde\A_0=\bzero_p$, and  $k=1,2,\dots.$  Note that for a terminal $k=K$, \eqref{eq:cuee} is equivalent to \eqref{eq:solution}.

For the variance of $\tilde{\bbeta}_{k}$, observe that
$$
\bzero = -M_{n_k,k}(\hat\bbeta_{n_k,k}) \approx -M_{n_k,k}(\check\bbeta_{n_k,k}) + \tilde\A_{n_{k},k} (\hat\bbeta_{n_k,k}-\check\bbeta_{n_k,k}).
$$
Thus, we have
$
\tilde\A_{n_{k},k} \check\bbeta_{n_k,k} + M_{n_k,k}(\check\bbeta_{n_k,k})  \approx  \tilde\A_{n_k,k} \hat\bbeta_{n_k,k}.
$
Using the above approximation, the variance formula is given by
\begin{align}
\nonumber
 \tilde \V_{k}  =& (\tilde{\A}_{k-1} + \tilde\A_{n_{k},k})^{-1} (
 \sum_{\ell=1}^k  \tilde\A_{n_\ell,\ell} \hat \V_{n_{\ell}, \ell} \tilde\A_{n_\ell,\ell}^{\top})
 [(\tilde\A_{k-1} + \tilde\A_{n_{k}, k})^{-1}]^{\top} \\
 \label{eq:Vtilde}
 = & (\tilde{\A}_{k-1} + \tilde\A_{n_{k},k})^{-1} (\tilde\A_{k-1} \tilde \V_{k-1}  \tilde\A_{k-1}^{\top} + \tilde\A_{n_{k},k} \hat \V_{n_{k}, k}\tilde\A_{n_{k}, k}^{\top})[(\tilde\A_{k-1} + \tilde\A_{n_{k}, k})^{-1}]^{\top},
\end{align}
for  $k=1,2,\dots$ and $\tilde\A_0=\tilde \V_0=\bzero_p$.

\begin{remark}
Under the normal linear regression model, all of the estimating equation estimators become ``exact'', in the sense that
 $
 \hat{\bbeta}_N=(\X'\X)^{-1}\X'\y=\hat{\bbeta}_{NK} = \hat{\bbeta}_{K} = \tilde{\bbeta}_{K}.
 $
\end{remark}

\subsection{Online Updating for Wald Tests}
Wald tests may be used to test individual coefficients or nested hypotheses based upon either the CEE or CUEE estimators from the cumulative data.  Let ($\breve{\bbeta}_k=(\breve{\beta}_{k,1},\ldots,\breve{\beta}_{k,p})',\breve\V_k$) refer to either the CEE regression coefficient estimator and corresponding variance in equations \eqref{eq:upbeta} and \eqref{eq:upvar}, or the CUEE regression coefficient estimator and corresponding variance in equations \eqref{eq:cuee} and \eqref{eq:Vtilde}.

To test $H_0: \beta_j=0$ at the $k^{th}$ update ($j=1,\ldots,p$), we may take the Wald statistic
$ z^{\ast 2}_{k,j} = \breve{\beta}_{k,j}^2/ \mbox{var}(\breve{\beta}_{k,j}), $
 or equivalently,
$ z^\ast_{k,j} = \breve{\beta}_{k,j}/ se(\breve{\beta}_{k,j}), $
where the standard error $se(\breve{\beta}_{k,j})=\sqrt{\mbox{var}(\breve{\beta}_{k,j})}$ and $\mbox{var}(\breve{\beta}_{k,j})$ is the $j^{th}$ diagonal element of $\breve\V_k.$  The corresponding p-value is $P(|Z| \geq |z^\ast_{k,j}|)=P(\chi^2_1 \geq z^{\ast 2}_{k,j})$ where $Z$ and $\chi^2_1$ are standard normal and 1 degree-of-freedom chi-squared random variables, respectively.

The Wald test statistic may also be used for assessing the difference between a full model M1 relative to a nested submodel M2. If $\bbeta$ is the parameter of model M1
and the nested submodel M2 is obtained from M1 by setting $\C\bbeta= \bzero$, where $\C$ is a rank $q$ contrast
matrix and $\breve \V$ is a consistent estimate of the covariance matrix of estimator $\breve\bbeta$, the test statistic is
$\breve\bbeta'\C'(\C\breve{\V}\C')^{-1}\C\breve\bbeta , $
which is distributed as $\chi^2_q$ under the null hypothesis that $\C\bbeta= \bzero.$ As an example, if M1 represents the full model containing all $p$ regression coefficients at the $k^{th}$ update, where the first coefficient $\beta_1$ is an intercept, we may test the global null hypothesis $H_0: \beta_2=\ldots=\beta_p=0$ with
$ w_k^\ast = \breve\bbeta_k'\C'(\C\breve{\V}_k\C')^{-1}\C\breve\bbeta_k , $
where $\C$ is $(p-1)\times p$ matrix $\C = [ \bzero , \I_{p-1} ]$ and the corresponding p-value is $P(\chi^2_{p-1} \geq w_k^\ast ).$

\subsection{Asymptotic Results}\label{sec:asymptotics}
In this section, we show consistency of the CUEE estimator.  Specifically, Theorem \ref{thm:cuee_consistent} shows that, under regularity, if the EE estimator based on the all $N$ observations $\hat\bbeta_N$ is a consistent estimator and the partition number $K$ goes to infinity, but not too fast, then the CUEE estimator $\tilde\bbeta_K$ is also a consistent estimator.  We first provide the technical regularity conditions.
We assume for simplicity of notation that $n_k=n$ for all $k=1,2,\ldots, K.$  Note that conditions (C1-C6) were given in \cite{Lin:Xi:aggr:2011}, and are provided below for completeness.  The proof of the theorem can be found in the Supplementary Material.

\begin{itemize}
\item[(C1)] The score function $\psi$ is measurable for any fixed $\bbeta$ and is twice continuously differentiable with respect to $\bbeta$.
\item[(C2)] The matrix $-\frac{\partial\psi(\z_i, \bbeta)}{\partial\bbeta}$ is semi-positive definite (s.p.d.), and $-\sum_{i=1}^n\frac{\partial\psi(\z_i, \bbeta)}{\partial\bbeta}$ is positive definite (p.d.) in a neighborhood of $\bbeta_0$ when $n$ is large enough.
\item[(C3)] The EE estimator $\hat{\bbeta}_{n,k}$ is strongly consistent, i.e. $\hat{\bbeta}_{n,k}\rightarrow \bbeta_0$ almost surely (a.s.) as $n \rightarrow \infty$.
\item[(C4)] There exists two p.d. matrices, $\bLambda_1$ and $\bLambda_2$ such that
$\bLambda_1 \leq n^{-1}\A_{n,k} \leq \bLambda_2$ for all $k = 1, \ldots, K,$ i.e. for any
$\bv\in \Real^p,$ $\bv'\bLambda_1\bv \leq n^{-1}\bv'\A_{n,k}\bv \leq \bv'\bLambda_1\bv$, where $\A_{n,k}$ is
given in \eqref{eq:A}.
\item[(C5)] In a neighborhood of $\bbeta_0$, the norm of the second-order derivatives $\frac{\partial^2\psi_j(\z_i, \bbeta)}{\partial\bbeta^2}$ is bounded uniformly, i.e. $\|\frac{\partial^2\psi_j(\z_i, \bbeta)}{\partial\bbeta^2}\| \le C_2$ for all $i, j,$ where $C_2$ is a constant.
\item[(C6)] There exists a real number $\alpha \in (1/4, 1/2)$ such that for any $\eta >0$, the EE estimator $\hat\bbeta_{n,k}$ satisfies $P(n^{\alpha}\|\hat\bbeta_{n,k}-\bbeta_{0}\| > \eta) \le C_{\eta}n^{2\alpha - 1}$, where $C_\eta>0$ is a constant only depending on $\eta$.
\end{itemize}

Rather than using condition (C4), we will use a slightly modified version which focuses on the behavior of $\A_{n,k}(\bbeta)$ for all $\bbeta$ in the neighborhood of $\bbeta_0$ (as in (C5)), rather than just at the subset estimate $\hat\bbeta_{n,k}.$
\begin{itemize}
\item[(C4')] In a neighborhood of $\bbeta_0$, there exists two p.d. matrices $\bLambda_1$ and $\bLambda_2$ such that $\bLambda_1 \le n^{-1}\A_{n,k}(\bbeta) \le \bLambda_2$ for all $\bbeta$ in the neighborhood of $\bbeta_0$ and for all $k=1, ..., K.$
\end{itemize}

\begin{thm}\label{thm:cuee_consistent}
Let $\hat\bbeta_N$ be the EE estimator based on entire data. Then under (C1)-(C2), (C4')-(C6), if the partition number $K$ satisfies $K=O(n^\gamma)$ for some $0<\gamma<min\{1-2\alpha, 4\alpha-1\}$, we have $P(\sqrt{N}\|\tilde\bbeta_{K}-\hat\bbeta_{N}\| > \delta)=o(1)$ for any $\delta>0.$
\end{thm}

\begin{remark}
If $n_k\neq n$ for all $k$, Theorem \ref{thm:cuee_consistent} will still hold, provided for each $k$, $\frac{n_{k-1}}{n_{k}}$ is bounded, where $n_{k-1}$ and $n_{k}$ are the respective sample sizes for subsets $k-1$ and $k$.
\end{remark}

\begin{remark}
Suppose $N$ independent observations $(y_i,\x_i),$ $i=1,\ldots,N$, where $y$ is a scalar response and $\x$ is a $p$-dimensional vector of predictor variables.  Further suppose $E(y_i) = g(\x_i'\bbeta)$ for $i=1,\ldots,N$ for $g$ a continuously differentiable function.
Under mild regularity conditions, \cite{Lin:Xi:aggr:2011} show in their Theorem 5.1 that condition (C6) is satisfied for 
a simplified version of the quasi-likelihood estimator of $\bbeta$ \citep{qle}, given as the solution to the estimating equation
$$Q(\bbeta)=\sum_{i=1}^N [y_i - g(\x_i'\bbeta)]\x_i=\bzero.$$
\end{remark}

\subsection{Rank Deficiencies in $\X_k$}\label{sec:rank}
Suppose $N$ independent observations $(y_i,\x_i),$ $i=1,\ldots,N$, where $y$ is a scalar response and $\x$ is a $p$-dimensional vector of predictor variables.  Using the same notation from the linear model setting, let $(y_{ki},\x_{ki}),$ $i = 1,\ldots,n_k$, be the observations from the $k^{th}$ subset where $\y_k=(y_{k1},y_{k2},\ldots, y_{kn_k})'$ and $\X_k = (\x_{k1},\x_{k2},\ldots,\x_{kn_k})'$.  For subsets $k$ in which $\X_k$ is not of full rank, we may have difficulty in solving the subset EE to obtain $\hat\bbeta_{n_k,k}$, which is used to compute both the AEE/CEE and CUEE estimators for $\bbeta$ in \eqref{eq:aee} and \eqref{eq:solution}, respectively.  However, just as in the linear model case, we can show under certain conditions that if $\X=(\X_1',\X_2',\dots,\X_K')'$ has full column rank $p$, then the estimators $\hat\bbeta_{NK}$ in \eqref{eq:aee} and $\tilde\bbeta_{K}$ in \eqref{eq:solution} for some terminal $K$ will be unique.  

Specifically, consider observations $(\y_k,\X_k)$ such that $E(y_{ki})=\mu_{ki} = g(\eta_{ki})$ with $\eta_{ki} = \x_{ki}'\bbeta$ for some known function $g$. The estimating function $\psi$ for the $k^{th}$ dataset is of the form
$$\psi(\z_{ki}, \bbeta) = \x_{ki} S_{ki} W_{ki}(y_{ki} - \mu_{ki}),~i=1,\ldots,n_k,$$
where $S_{ki} = \partial \mu_{ki} / \partial \eta_{ki}$, and
$W_{ki}$ is a positive and possibly data dependent weight. Specifically, $W_{ki}$ may depend on $\bbeta$ only through $\eta_{ki}.$ In matrix form, the estimating equation becomes
\begin{eqnarray}\label{eq:eefun}
\X_k' \bS'_k \W_k (\y_k-\bmu_k) &=& \bzero,
\end{eqnarray}
where $\bS_k = \mbox{Diag}(S_{k1}, \ldots, S_{kn_k})$,
$\W_k = \mbox{Diag}(W_{k1}, \ldots, W_{kn_k})$, 
and $\bmu_k = \big(\mu_{k1}, \ldots, \mu_{kn_k}\big)'$.

With $\bS_k$, $\W_k$, and $\bmu_k$ evaluated at some initial value $\bbeta^{(0)}$, the standard Newton--Raphson method for the iterative solution of \eqref{eq:eefun} solves the linear equations
\begin{equation}
\label{eq:eeup}
\X_k' \bS'_k \W_k \bS_k \X_k (\bbeta - \bbeta^{(0)}) = \X_k' \bS_k' \W_k (\y_k - \bmu_k)
\end{equation}
for an updated $\bbeta$.

Rewrite equation~\eqref{eq:eeup} as
\begin{equation}
  \label{eq:irls}
  \X_k' \bS'_k \W_k \bS_k \X_k \bbeta = \X_k' \bS'_k \W_k \bv_k
\end{equation}
where  $\bv_k = \y_k - \bmu_k  +  \bS_k \X_k \bbeta^{(0)}$.
Equation \eqref{eq:irls} is the normal equation of a weighted least squares regression
with response $\bv_k$, design matrix $\bS_k \X_k$, and weight $\W_k$.
Therefore the iterative reweighted least squares approach (IRLS)
can be used to implement the Newton--Raphson method for an iterative solution to \eqref{eq:eefun} \citep[e.g.,][]{green:1984}.

Rank deficiency in $\X_k$ calls for a generalized inverse of
$\X_k' \bS'_k \W_k \bS_k \X_k$. In order to show uniqueness of estimators $\hat\bbeta_{NK}$ in \eqref{eq:aee} and $\tilde\bbeta_{K}$ in \eqref{eq:solution} for some terminal $K$, we must first establish that the IRLS algorithm will work and converge for subset $k$ given the same initial value $\bbeta^{(0)}$ when $\X_k$ is not of full rank. Upon convergence of IRLS at subset $k$ with solution $\hat\bbeta_{n_k,k}$, we must then verify that the CEE and CUEE estimators that rely on $\hat\bbeta_{n_k,k}$ are unique.  The following proposition summarizes the result; the proof is provided in the Supplementary Material.

\begin{prop}\label{prop:glm}
Under the above formulation, assuming that conditions (C1-C3) hold for a full-rank sub-column matrix of $\X_k$, estimators $\hat\bbeta_{NK}$ in \eqref{eq:aee} and $\tilde\bbeta_{K}$ in \eqref{eq:solution} for some terminal $K$ will be unique provided $\X$ is of full rank.
\end{prop}

The simulations in Section \ref{sec:sim.ee} consider rank deficiencies in binary logistic regression and Poisson regression.
Note that for these models, the variance of the estimators $\hat\bbeta_K$ and $\tilde\bbeta_K$ are given by $\A_K^{-1}=(\sum_{k=1}^K\A_{n_k,k})^{-1}$ or  $\tilde\A_K^{-1}=(\sum_{k=1}^K\tilde\A_{n_k,k})^{-1}.$  For robust sandwich estimators, for those subsets $k$ in which $\A_{n_k,k}$ is not invertible, we replace $\A_{n_k,k} \hat\V_{n_k,k}  \A_{n_k,k}^{\top}$ and $\tilde\A_{n_k,k} \hat \V_{n_k,k}  \tilde\A_{n_k,k}^{\top}$ in the ``meat'' of equations \eqref{eq:upvar} and \eqref{eq:Vtilde}, respectively, with an estimate of $\Q_{n_k,k}$ from \eqref{eq:aeevar}.  In particular, we use $\hat \Q_{n_k,k}=\sum_{i=1}^{n_k} \psi(\z_{ki}, \hat\bbeta_k)\psi(\z_{ki}, \hat\bbeta_k)^{\top}$ for the CEE variance and $\tilde \Q_{n_k,k}=\sum_{i=1}^{n_k} \psi(\z_{ki}, \tilde\bbeta_k)\psi(\z_{ki}, \tilde\bbeta_k)^{\top}$ for the CUEE variance. We use these modifications in the robust Poisson regression simulations in Section \ref{sec:sim.glm} for the CEE and CUEE estimators, as by design, we include binary covariates with somewhat low success probabilities.  Consequently, not all subsets $k$ will observe both successes and failures, particularly for covariates with success probabilities of 0.1 or 0.01, and the corresponding design matrices $\X_k$ will not always be of full rank. Thus $\A_{n_k,k}$ will not always be invertible for finite $n_k$, but will be invertible for large enough $n_k$.  We also perform proof of concept simulations in Section \ref{sec:sim.glm.ginv} in binary logistic regression, where we compare CUEE estimators under different choices of generalized inverses.

    
\section{Simulations}\label{sec:sim}
\subsection{Normal Linear Regression: Residual Diagnostic Performance}

In this section we evaluate the performance of the outlier tests discussed in Section \ref{sec:outlier.test}.  Let $k^\ast$ denote the index of the single subset of data containing any outliers. We generated the data according to the model
\begin{equation}\label{eq:outlier.gen}
y_{k i}=\x_{k i}'\bbeta + \epsilon_{k i} + b_k\delta\eta_{k i},~~~i=1,\ldots, n_k,
\end{equation}
where $b_k=0$ if $k\ne k^\ast$ and $b_k\sim \mbox{Bernoulli}(0.05)$ otherwise. Notice that the first two terms on the right-hand-side correspond to the usual linear model with
 $\bbeta=(1,2,3,4,5)'$, $x_{k i[2:5]} \sim N(\bzero,\I_4)$ independently, $x_{k i[1]} =1$,  and
 $\epsilon_{k i}$ are the independent errors, while the final term is responsible for generating the outliers.  Here,  $\eta_{k i} \sim \mbox{Exp}(1)$ independently and $\delta$ is the scale parameter controlling magnitude or strength of the outliers.  We set $\delta \in \{0,2,4,6\}$ corresponding to ``no'', ``small'', ``medium'', and ``large'' outliers.

To evaluate the performance of the individual outlier test in \eqref{eq:t.test}, we generated the random errors as $\epsilon_{k i}\sim \mbox{N}(0,1).$  To evaluate the performance of the global outlier tests in \eqref{eq:f.test} and \eqref{eq:f.test.asymp}, we additionally considered $\epsilon_{k i}$ as independent skew-t variates with degrees of freedom $\nu=3$ and skewing parameter $\gamma=1.5$, standardized to have mean 0 and variance 1.  
To be precise, we use the skew $t$ density
\begin{equation*}
g(x) = \left\{\begin{array}{l}\frac{2}{\gamma + \frac{1}{\gamma}}f(\gamma x) \text{\quad for $x<0$}\\
\frac{2}{\gamma + \frac{1}{\gamma}}f(\frac{x}{\gamma}) \text{\quad for $x\ge 0$}\end{array}\right.
\end{equation*}
where $f(x)$ is the density of the $t$ distribution with $\nu$ degrees of freedom.

For all outlier simulations, we varied $k^\ast$, the location along the data stream in which the outliers occur.  We also varied $n_k=n_k\ast\in\{100,500\}$ which additionally controls the number of outliers in dataset $k^\ast.$  For each subset $\ell=1,\ldots, k^\ast-1$ and for 95\% of observations in subset $k^\ast$, the data did not contain any other outliers.

\begin{table}[!t]
	{\singlespace
	\caption{Power of the outlier tests for various locations of outliers ($k^\ast$), subset sample sizes ($n_k=n_{k\ast}$), and outlier strengths (no, small, medium, large).  Within each cell, the top entry corresponds to the normal-based $F$ test and the bottom entry corresponds to the asymptotic $F$ test that does not rely on normality of the errors.}
	\label{tab:Fpower}
	\centering
    \footnotesize
\begin{tabular}{lcccccccr}
\hline
Outlier   & \multicolumn{4}{c}{$n_{k\ast}=100$ (5 true outliers) } & \multicolumn{4}{c}{$n_{k\ast}=500$ (25 true outliers)}\\
Strength  &  $k^\ast=5$ &  $k^\ast=10$  & $k^\ast=25$  & $k^\ast=100$&  $k^\ast=5$ & $k^\ast=10$  & $k^\ast=25$ & $k^\ast=100$   \\ \hline
& \multicolumn{4}{c}{$F$ Test/Asymptotic $F$ Test(m=2)}&\multicolumn{4}{c}{$F$ Test/Asymptotic $F$ Test(m=2)}\\
\hline
\multicolumn{6}{l}{\underline{Standard Normal Errors}} \\
\rule{0pt}{3ex} none   &0.0626 &0.0596 &0.0524 &0.0438  &0.0580  &0.0442 &0.0508 &0.0538\\
                       &0.0526 &0.0526 &0.0492 &0.0528  &0.0490  &0.0450 &0.0488 &0.0552\\
\rule{0pt}{3ex} small  &0.5500 &0.5690 &0.5798 &0.5718  &0.9510  &0.9630 &0.9726 &0.9710\\
                       &0.2162 &0.2404 &0.2650 &0.2578  &0.6904  &0.7484 &0.7756 &0.7726\\
\rule{0pt}{3ex} medium &0.9000 &0.8982 &0.9094 &0.9152  &1.0000  &1.0000 &1.0000 &1.0000\\
                       &0.5812 &0.6048 &0.6152 &0.6304  &0.9904  &0.9952 &0.9930 &0.9964\\
\rule{0pt}{3ex} large  &0.9680 &0.9746 &0.9764 &0.9726  &1.0000  &1.0000 &1.0000 &1.0000\\
                       &0.5812 &0.6048 &0.6152 &0.6304  &0.9998  &1.0000 &1.0000 &1.0000\\ \hline
\multicolumn{6}{l}{\underline{Standardized Skew t Errors}} \\
\rule{0pt}{3ex} none   &0.2400 &0.2040 &0.1922 &0.1656  &0.2830 &0.2552 &0.2454 &0.2058\\
                       &0.0702 &0.0630 &0.0566 &0.0580  &0.0644 &0.0580 &0.0556 &0.0500\\
\rule{0pt}{3ex} small  &0.5252 &0.4996 &0.4766 &0.4520  &0.7678 &0.7598 &0.7664 &0.7598\\
                       &0.2418 &0.2552 &0.2416 &0.2520  &0.6962 &0.7400 &0.7720 &0.7716\\
\rule{0pt}{3ex} medium &0.8302 &0.8280 &0.8232 &0.8232  &0.9816 &0.9866 &0.9928 &0.9932\\
                       &0.5746 &0.5922 &0.6102 &0.6134  &0.9860 &0.9946 &0.9966 &0.9960\\
\rule{0pt}{3ex} large  &0.9296 &0.9362 &0.9362 &0.9376  &0.9972 &0.9970 &0.9978 &0.9990\\
                       &0.7838 &0.8176 &0.8316 &0.8222  &0.9988 &0.9992 &0.9998 &1.0000\\ \hline
\end{tabular}

  \vspace*{0.1in}

 \hspace*{0.5in} Power with ``outlier strength $=$ no" are Type I errors.
}
\end{table}


To evaluate the global outlier tests \eqref{eq:f.test} and \eqref{eq:f.test.asymp} with $m=2$, we estimated power using $B=500$ simulated data sets with significance level $\alpha=0.05,$ where power was estimated as the proportion of 500 datasets in which $\check{F}_{k^\ast} \geq F(0.95,n_{k^\ast},N_{{k^\ast}-1}-5)$ or $\check{F}^a_{k^\ast} \geq F(0.95, 2, N_{k^\ast-1}-1).$ The power estimates for the various subset sample sizes $n_{k^\ast},$ locations of outliers $k^\ast,$ and outlier strengths $\delta$ appear in Table \ref{tab:Fpower}. When the errors were normally distributed (top portion of table), notice that the Type I error rate was controlled in all scenarios for both the $F$ test and asymptotic $F$ test.  As expected, power tends to increase as outlier strength and/or the number of outliers increase.  Furthermore, larger values of $k^\ast,$ and hence greater proportions of ``good'' outlier-free  data, also tend to have higher power; however, the magnitude of improvement decreases once the denominator degrees of freedom ($N_{k^\ast-1}-p$ or $N_{k^\ast-1}-m+1$)  become large enough, and the $F$ tests essentially reduce to $\chi^2$ tests. Also as expected, the $F$ test given by \eqref{eq:f.test} is more powerful than the asymptotic $F$ test given in \eqref{eq:f.test.asymp} when, in fact, the errors were normally distributed. When the errors were not normally distributed (bottom portion of table), the empirical type I error rates of the $F$ test given by \eqref{eq:f.test} are severely inflated and hence, its empirical power in the presence of outliers cannot be trusted.  The asymptotic $F$ test, however, maintains the appropriate size.

\begin{figure}[!t]
	\centering
	{\singlespace
		\includegraphics[width=\textwidth]{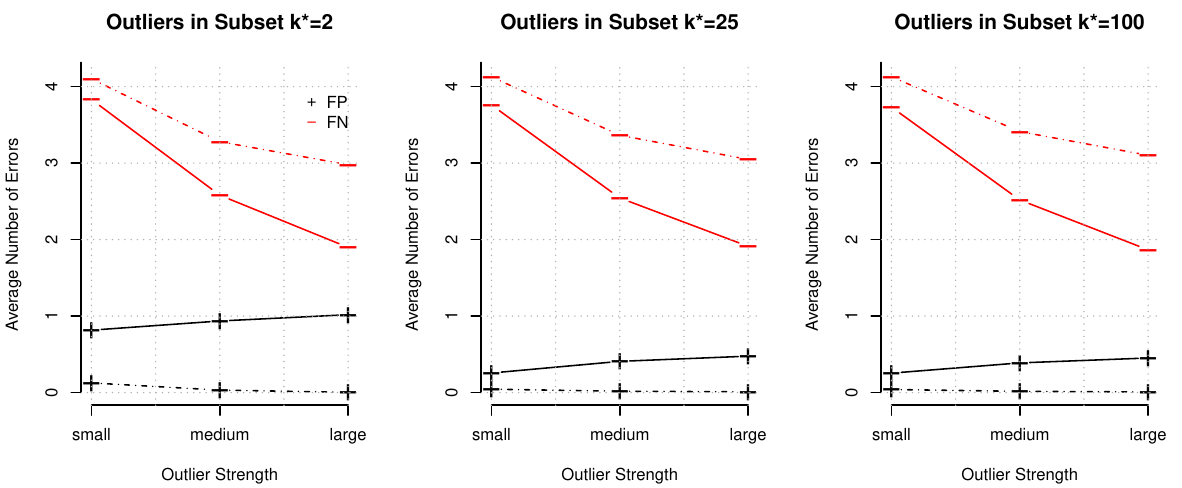}
		\includegraphics[width=\textwidth]{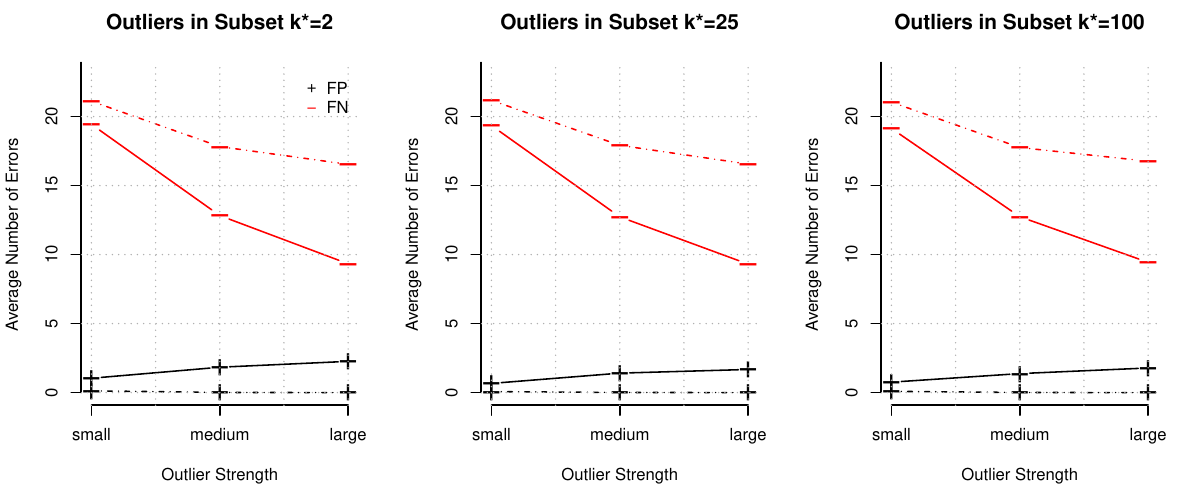}
	\caption{Average numbers of False Positives and False Negatives for outlier t-tests for $n_{k^\ast}=100$ (top) and $n_{k^\ast}=500$ (bottom). Solid lines correspond to the predictive residual test while dotted lines correspond to the externally studentized residuals test using only data from subset $k^\ast$.}
	\label{fig:ttest.fpfn.n100n500}}
\end{figure}

For the outlier $t$-test in \eqref{eq:t.test}, we examined the average number of false negatives (FN) and average number of false positives (FP) across the $B=500$ simulations. False negatives and false positives were declared based on a Benjamini-Hochberg adjusted $p$-value threshold of 0.10.  These values were plotted in solid lines against outlier strength in Figure \ref{fig:ttest.fpfn.n100n500} for $n_{k^\ast}=100$ and $n_{k^\ast}=500$ for various values of $k^\ast$ and $\delta.$  Within each plot the FN decreases as outlier strength increases, and also tends to decrease slightly across the plots as $k^\ast$ increases.  FP increases slightly as outlier strength increases, but decreases as $k^\ast$ increases.  As with the outlier $F$ test, once the degrees of freedom $N_{k^\ast-1}-p$ get large enough, the $t$-test behaves more like a $z$-test based on the standard normal distribution.  For comparison, we also considered FN and FP for an outlier test based upon the externally studentized residuals from subset $k^\ast$ only.  Specifically, under model \eqref{eq:klinmod}, the externally studentized residuals $t_{k^\ast i}$ as given by \eqref{eq:indiv.res} follow a $t$ distribution with $n_{k^\ast}-p-1$ degrees of freedom.  Again, false negatives and false positives were declared based on a Benjamini-Hochberg adjusted $p$-value threshold of 0.10, and the FN and FP for the externally studentized residual test are plotted in dashed lines in Figure \ref{fig:ttest.fpfn.n100n500} for $n_{k^\ast}=100$ and $n_{k^\ast}=500$. This externally studentized residual test tends to have a lower FP, but higher FN than the predictive residual test that uses the previous data.  Also, the FN and FP for the externally studentized residual test are essentially constant across $k^\ast$ for fixed $n_{k^\ast},$ as the externally studentized residual test relies on only the current dataset of size $n_{k^\ast}$ and not the amount of previous data controlled by $k^\ast.$ Consequently, the predictive residual test has improved power over the externally studentized residual test, while still maintaining a low number of FP.  Note that the average false discovery rate for the predictive residual test based on Benjamini-Hochberg adjusted $p$-values was controlled in all cases except when $k^\ast=2$ and $n_{k\ast}=100,$ representing the smallest sample size considered.

\subsection{Simulations for Estimating Equations}\label{sec:sim.ee}

\begin{figure}[!t]
	\centering
	{\singlespace
		\includegraphics[width=\textwidth]{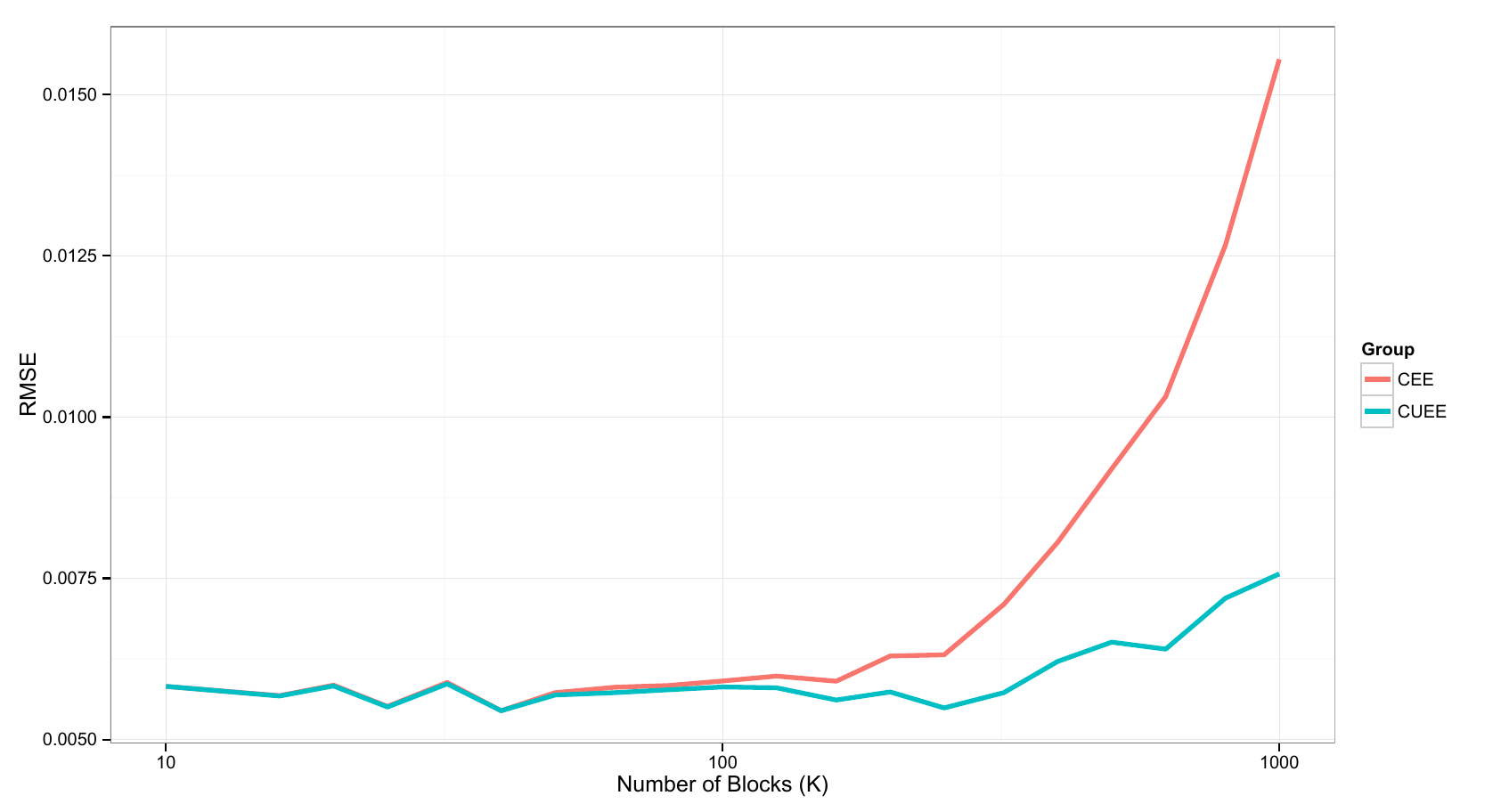}
	\caption{RMSE comparison between CEE and CUEE estimators for different numbers of blocks.}
	\label{fig:rmse}}
\end{figure}

\subsubsection{Logistic Regression}
To examine the effect of the total number of blocks $K$ on the performance of the CEE and CUEE estimators, we generated $y_i\sim \mbox{Bernoulli}(\mu_i),$ independently for $i=1,\ldots,100000,$ with
$ \mbox{logit}(\mu_i) = \x_i'\bbeta $
where
$\bbeta=(1,1,1,1,1,1)'$, $x_{i[2:4]}\sim \mbox{Bernoulli}(0.5)$ independently, $x_{i[5:6]}\sim N(\bzero,\I_2)$ independently, and $x_{k i[1]} =1$. The total sample size was fixed at $N = 100000$, but in computing the CEE and CUEE estimates, the number of blocks $K$ varied from 10 to 1000 where $N$ could be divided evenly by $K$. At each value of $K$, the root-mean square error (RMSE) of both the CEE and CUEE estimators were calculated as $\sqrt{\frac{\sum_{j=1}^6(\breve{\beta}_{Kj} -1)^2}{6}},$ where $\breve{\beta}_{Kj}$ represents the $j^{th}$ coefficient in either the CEE or CUEE terminal estimate. The averaged RMSEs are obtained with 200 replicates. Figure \ref{fig:rmse} shows the plot of averaged RMSEs versus the number of blocks $K$. It is obvious that as the number of blocks increases (block size decreases), RMSE from CEE method increases very fast while RMSE from the CUEE method remains relatively stable.

\subsubsection{Robust Poisson Regression}\label{sec:sim.glm}

\begin{figure}[!t]
	\centering
	{\singlespace
		\includegraphics[width=\textwidth]{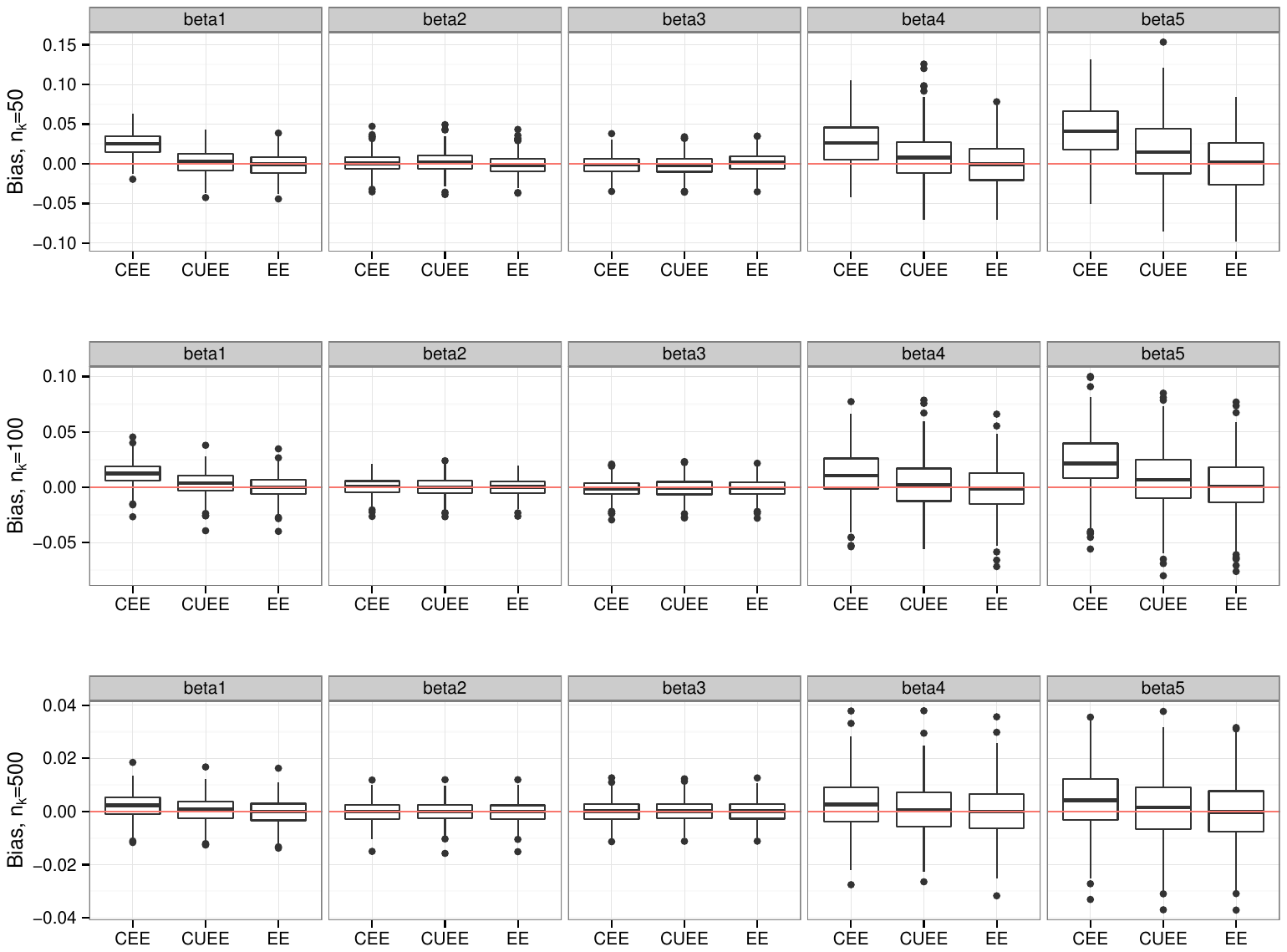}
	\caption{Boxplots of biases for 3 types of estimators (CEE, CUEE, EE) of $\beta_j$ (estimated $\beta_j$ - true $\beta_j$), $j=1,\ldots,5$, for varying $n_k.$}
	\label{fig:boxbias}}
\end{figure}

In these simulations, we compared the performance of the (terminal) CEE and CUEE estimators with the EE estimator based on all of the data.
We generated $B=500$ datasets of $y_i\sim \mbox{Poisson}(\mu_i),$ independently for $i=1,\ldots,N$ with
$ \log(\mu_i) = \x_i'\bbeta $
where
$\bbeta=(0.3,-0.3,0.3,-0.3,0.3)'$, $x_{k i[1]} =1$, $x_{i[2:3]} \sim N(\bzero,\I_2)$ independently, $x_{i[4]} \sim Bernoulli(0.25)$ independently, and $x_{i[5]} \sim Bernoulli(0.1)$ independently. We fixed $K=100,$ but varied $n_k=n\in\{50,100,500\}.$

\begin{figure}[!t]
	\centering
	{\singlespace
		\includegraphics[width=\textwidth]{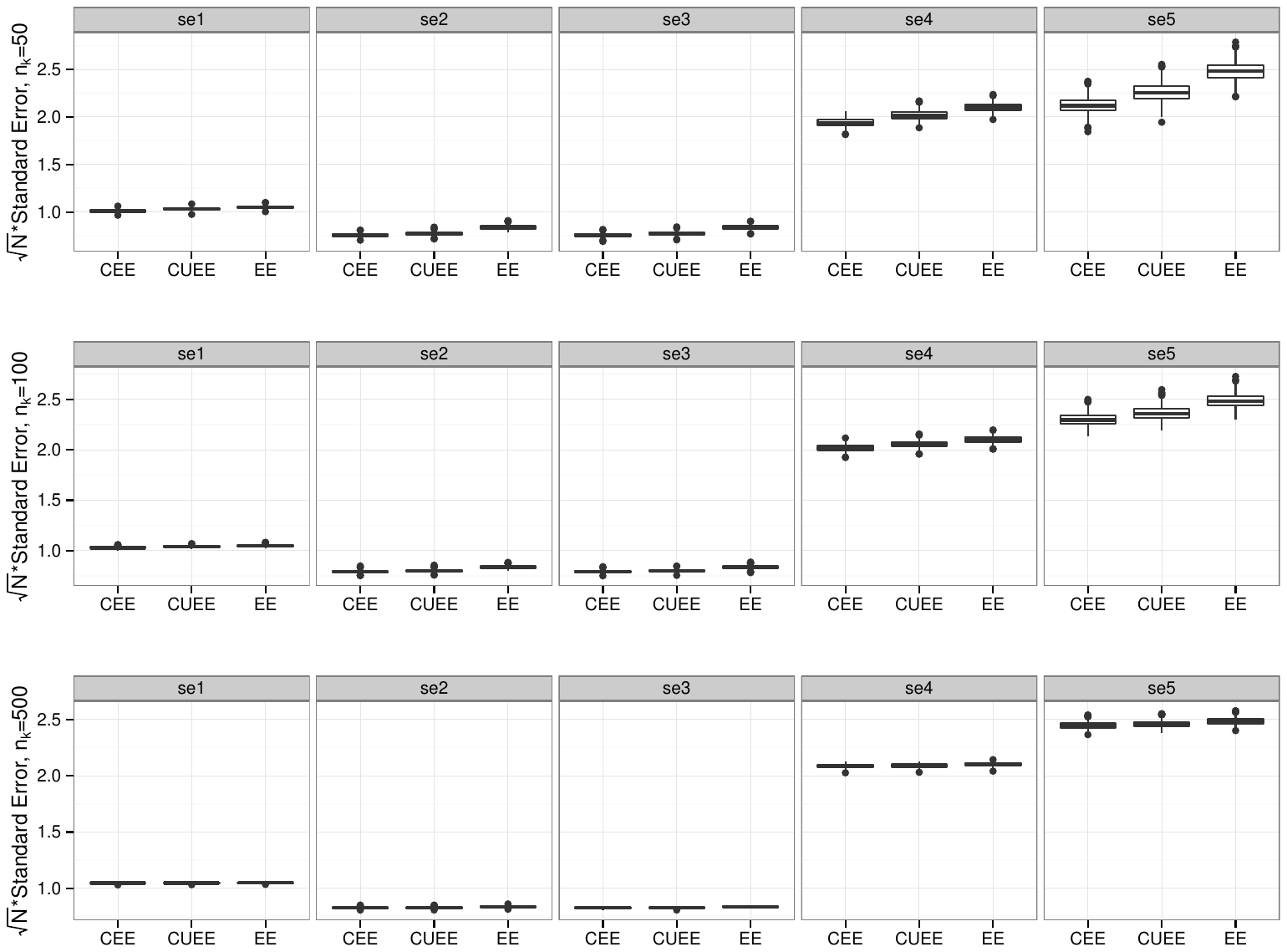}
	\caption{Boxplots of standard errors for 3 types of estimators (CEE, CUEE, EE) of $\beta_j$, $j=1,\ldots,5,$ for varying $n_k.$ Standard errors have been multiplied by $\sqrt{Kn_k}=\sqrt{N}$ for comparability.}
	\label{fig:boxse}}
\end{figure}

Figure \ref{fig:boxbias} shows boxplots of the biases in the 3 types of estimators (CEE, CUEE, EE) of $\beta_j,$ $j=1,\ldots,5,$ for varying $n_k.$  The CEE estimator tends to be the most biased, particularly in the intercept, but also in the coefficients corresponding to binary covariates. The CUEE estimator also suffers from slight bias, while the EE estimator performs quite well, as expected.  Also as expected, as $n_k$ increases, bias decreases.  The corresponding robust (sandwich-based) standard errors are shown in Figure \ref{fig:boxse}, but the results were very similar for variances estimated by $\A_K^{-1}$ and $\tilde\A_K^{-1}$.  In the plot, as $n_k$ increases, the standard errors become quite similar for the three methods.

Table \ref{tab:rmseratio} shows the coefficient-wise RMSE ratios :
$$ \frac{\mbox{RMSE(CEE)}}{\mbox{RMSE(EE)}} \mbox{~~and~~} \frac{\mbox{RMSE(CUEE)}}{\mbox{RMSE(EE)}},$$ where we take the RMSE of the EE estimator as the gold standard.
The RMSE ratios for CEE and CUEE estimators confirm the boxplot results in that the intercept and the coefficients corresponding to binary covariates ($\beta_4$ and $\beta_5$) tend to be the most problematic for both estimators, but more so for the CEE estimator.

\begin{figure}[!t]
	\centering
	{\singlespace
		\includegraphics[width=\textwidth]{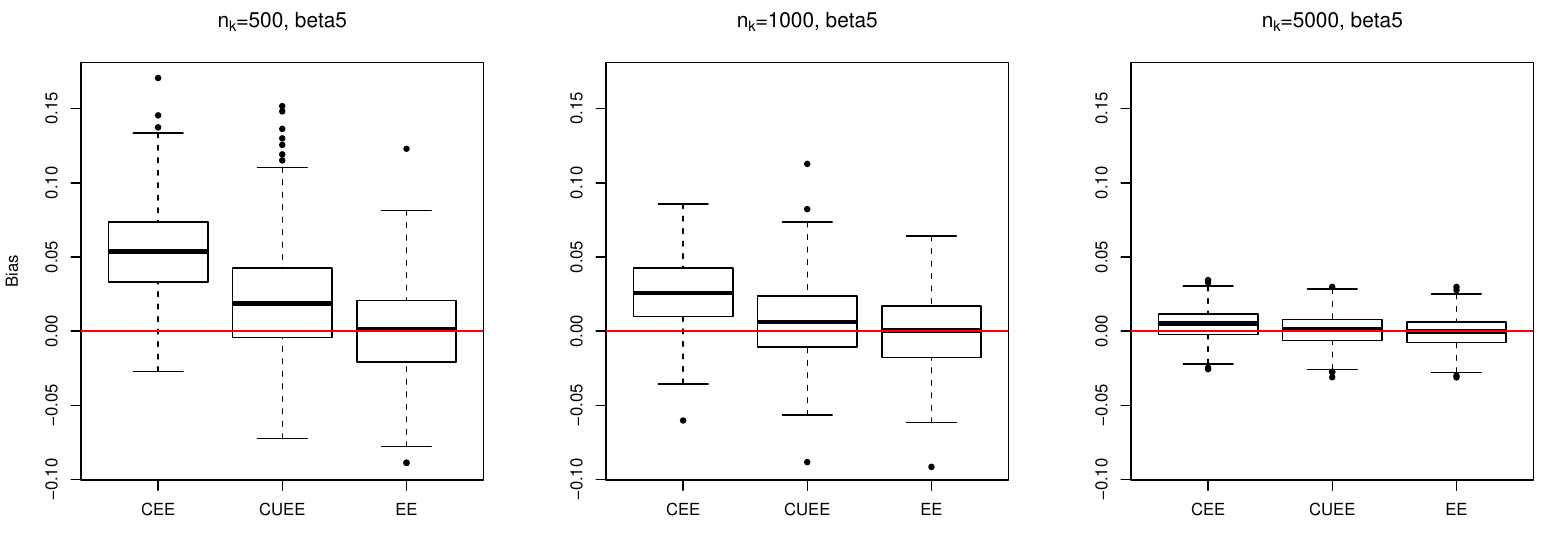}
	\caption{Boxplots of biases for 3 types of estimators (CEE, CUEE, EE) of $\beta_5$ (estimated $\beta_5$ - true $\beta_5$), for varying $n_k,$ when $x_{i[5]} \sim Bernoulli(0.01)$.}
	\label{fig:boxbiasbeta5}}
\end{figure}

For this particular simulation, it appears $n_k=500$ is sufficient to adequately reduce the bias.  However, the appropriate subset size $n_k$, if given the choice, is relative to the data at hand.  For example, if we alter the data generation of the simulation to instead have $x_{i[5]} \sim Bernoulli(0.01)$ independently, but keep all other simulation parameters the same, the bias, particularly for $\beta_5$, still exists at $n_k=500$ (see Figure \ref{fig:boxbiasbeta5}) but diminishes substantially with $n_k=5000.$

\begin{table}[!t]
	\centering
	{\singlespace
		\caption{Root Mean Square Error Ratios of CEE and CUEE with EE}
	\label{tab:rmseratio}
\begin{tabular}{lrrrrrr}
\hline
& & $\beta_1$ & $\beta_2$ & $\beta_3$ & $\beta_4$ & $\beta_5$  \\
\hline
\rule{0pt}{2ex}$n_k=50$ & CEE  & 4.133 & 1.005 & 1.004 & 1.880 & 2.288 \\
                        & CUEE & 1.180 & 1.130 & 1.196 & 1.308 & 1.403 \\
\rule{0pt}{2ex}$n_k=100$ & CEE &  2.414 & 1.029 & 1.036 & 1.299 & 1.810 \\
                        & CUEE & 1.172 & 1.092 & 1.088 & 1.118 & 1.205 \\
\rule{0pt}{2ex}$n_k=500$ & CEE & 1.225 & 1.002 & 1.002 & 1.060 & 1.146 \\
                         & CUEE & 0.999 & 1.010 & 1.016 & 0.993 & 1.057 \\
\hline
\end{tabular}}
\end{table}

\subsubsection{Rank Deficiency and Generalized Inverse }\label{sec:sim.glm.ginv}

Consider the CUEE estimator for a given dataset under two choices of generalized inverse, the Moore-Penrose generalized inverse, and a generalized inverse generated according to Theorem 2.1 of \cite{rao:1972}.  For this small-scale, proof-of-concept simulation, we generated $B=100$ datasets of $y_i\sim \mbox{Bernoulli}(\mu_i),$ independently for $i=1,\ldots,20000,$ with
$ \mbox{logit}(\mu_i) = \x_i'\bbeta $
where
$\bbeta=(1,1,1,1,1)'$, $x_{i[2]}\sim \mbox{Bernoulli}(0.5)$ independently, $x_{i[3:5]}\sim N(\bzero,\I_3)$ independently, and $x_{k i[1]} =1$. We fixed $K=10$  and $n_k=2000.$  The pairs of $(y_i,\x_i)$ observations were considered in different orders, so that in the first ordering all subsets would result in $\A_{n_k,k}$ being of full rank, $k=1,\ldots, K,$ but in the second ordering all of the subsets would not have full rank $\A_{n_k,k}$ due to the grouping of the zeros and ones from the binary covariate.  In the first ordering, we used the initially proposed CUEE estimator $\tilde\bbeta_K$ in \eqref{eq:solution} to estimate $\bbeta$ and its corresponding variance $\tilde \V_K$ in \eqref{eq:Vtilde}.  In the second ordering, we used two different generalized inverses to compute $\hat\bbeta_{n_k,k},$ denoted by CUEE$^{(-)}_1$ and CUEE$^{(-)}_2$ in Table \ref{tab:ginv}, with variance given by $\tilde\A_K^{-1}$.  The estimates reported in Table \ref{tab:ginv} were averaged over 100 replicates. The corresponding EE estimates, which are computed by fitting all $N$ observations simultaneously, are also provided for comparison.  As expected, the values reported for CUEE$^{(-)}_1$ and CUEE$^{(-)}_2$ are identical, indicating that the estimator is invariant to the choice of generalized inverse, and these results are quite similar to those of the EE estimator and CUEE estimator with all full-rank matrices $\A_{n_k,k}$, $k=1,\ldots,K$.

\section{Data Analysis}\label{sec:data}

We examined the airline on-time statistics, available at \url{http://stat-computing.org/dataexpo/2009/the-data.html}.
The data consists of flight arrival and departure details for all
commercial flights within the USA, from October 1987 to April
2008.  This involves $N=123,534,969$ observations and 29 variables ($\sim 11$ GB).

We first used logistic regression to model the probability of late arrival (binary; 1 if late by more than 15 minutes, 0 otherwise) as a function of departure time (continuous);
 distance (continuous, in thousands of miles), day/night flight status (binary; 1 if departure between 8pm and 5am, 0 otherwise);
 weekend/weekday status (binary; 1 if departure occurred during the weekend, 0 otherwise), and distance type (categorical; `typical distance' for distances less than 4200 miles, the reference level `large distance' for distances between 4200 and 4300 miles, and `extreme distance' for distances greater than 4300 miles) for $N=120,748,239$ observations with complete data.  

\begin{table}[!t]
	\centering
		{\singlespace
		\caption{Estimates and standard errors for CUEE$^{(-)}_1$, CUEE$^{(-)}_2$, CUEE, and EE estimators. CUEE$^{(-)}_1$ and CUEE$^{(-)}_2$ correspond to CUEE estimators using two different generalized inverses for $\A_{n_k,k}$ when $\A_{n_k,k}$ is not invertible.}}
	\label{tab:ginv}
	{\singlespace
	{\footnotesize
\begin{tabular}{@{}rr|rr|rr|rr@{}}\hline\hline
            \multicolumn{2}{c|}{CUEE$^{(-)}_1$} & \multicolumn{2}{c|}{CUEE$^{(-)}_2$} & \multicolumn{2}{c|}{CUEE} & \multicolumn{2}{c}{EE}\\
\rule{0pt}{3ex}  $\tilde\bbeta_{Kj}$ & $se(\tilde\bbeta_{Kj})$ & $\tilde\bbeta_{Kj}$ & $se(\tilde\bbeta_{Kj})$ & $\tilde\bbeta_{Kj}$ & $se(\tilde\bbeta_{Kj})$  & $\hat\bbeta_{Nj}$ & $se(\hat\bbeta_{Nj})$ \\
 \hline
\rule{0pt}{3ex} 0.9935731  &0.02850429  &0.9935731 & 0.02850429 &0.9940272    & 0.02847887 &0.9951570 & 0.02845648\\
\rule{0pt}{3ex} 0.8902375  &0.03970919  &0.8902375 & 0.03970919 &0.8923991    & 0.03936931 &0.8933344 & 0.03935490\\
\rule{0pt}{3ex} 0.9872035  &0.02256396  &0.9872035 & 0.02256396 &0.9879017    & 0.02247598 &0.9891857 & 0.02245082\\
\rule{0pt}{3ex} 0.9916863  &0.02264102  &0.9916863 & 0.02264102 &0.9925716    & 0.02248187 &0.9938864 & 0.02246949\\
\rule{0pt}{3ex} 0.9874042  &0.02260353  &0.9874042 & 0.02260353 &0.9882167    & 0.02247671 &0.9895110 & 0.02244759\\
\hline\hline
\end{tabular}}}
\end{table}

For CEE and CUEE, we used a subset size of $n_k=50,000$ for $k=1,\ldots,K-1$, and $n_K=48239$ to estimate the data in the online-updating framework. However, to avoid potential data separation problems due to rare events (extreme distance; 0.021\% of the data with 26,021 observations), 
a detection mechanism has been introduced at each block. If such a problem exists, the next block of data will be combined until the problem disappears. We also computed EE estimates and standard errors using the commercial software Revolution R.

All three methods agree that all covariates except extreme distance are highly associated with late flight arrival ($p<0.00001$), with later departure times and longer distances corresponding to a higher likelihood for late arrival, and night-time and weekend flights corresponding to a lower likelihood for late flight arrival (see Table \ref{tab:airline}). However, extreme distance is not associated with the late flight arrival ($p = 0.613$). 
The large $p$ value also indicates that even if number of observations is huge, there is no guarantee that all covariates must be significant.
As we do not know the truth in this real data example, we compare the estimates and standard errors of CEE and CUEE with those from Revolution R, which computes the EE estimates, but notably not in an online-updating framework. In Table \ref{tab:airline}, the CUEE and Revolution R regression coefficients tend to be the most similar. 
The regression coefficient estimates and standard errors for CEE are also close to those from Revolution R, with the most discrepancy in the regression coefficients again appearing in the intercept and coefficients corresponding to binary covariates.

\begin{table}[!t]
	\centering
	{\singlespace
	\caption{Estimates and standard errors ($\times 10^5$) from the Airline On-Time data for EE (computed by Revolution R), CEE, and CUEE estimators.}
	\label{tab:airline}
\begin{tabular}{@{}l|rr|rr|rr@{}}\hline\hline
            & \multicolumn{2}{c|}{EE} & \multicolumn{2}{c|}{CEE}  &\multicolumn{2}{c}{CUEE}\\
\rule{0pt}{2ex} & $\hat\bbeta_{Nj}$ & $se(\hat\bbeta_{Nj})$ & $\hat\bbeta_{Kj}$ & $se(\hat\bbeta_{Kj})$ & $\tilde\bbeta_{Kj}$ & $se(\tilde\bbeta_{Kj})$  \\
 \hline
\rule{0pt}{2ex}Intercept & $-3.8680367 $ & 1395.65 & $-3.70599823$ & $ 1434.60 $& $-3.880106804$ & $1403.49$ \\
\rule{0pt}{2ex}Depart    & $ 0.1040230 $ &    6.01 & $ 0.10238977$ & $    6.02 $& $ 0.101738105$ & $   5.70$ \\
\rule{0pt}{2ex}Distance  & $ 0.2408689 $ &   40.89 & $ 0.23739029$ & $   41.44 $& $ 0.252600016$ & $  38.98$ \\
\rule{0pt}{2ex}Night     & $-0.4483780 $ &   81.74 & $-0.43175229$ & $   82.15 $& $-0.433523534$ & $  80.72$ \\
\rule{0pt}{2ex}Weekend   & $-0.1769347 $ &   54.13 & $-0.16943755$ & $   54.62 $& $-0.177895116$ & $  53.95$ \\
\rule{0pt}{2ex}TypDist   & $ 0.8784740 $ & 1389.11 & $ 0.76748539$ & $ 1428.26 $& $ 0.923077960$ & $1397.46$ \\
\rule{0pt}{2ex}ExDist     & $-0.0103365 $ & 2045.71 & $-0.04045108$ & $ 2114.17 $& $-0.009317274$ & $2073.99$ \\

\hline\hline
\end{tabular}}
\end{table}

We finally considered arrival delay ($ArrDelay$) as a continuous variable by modeling $\log(ArrDelay-min(ArrDelay)+1)$ as a function of departure time, distance, day/night flight status, and weekend/weekday flight status for United Airline flights ($N=13,299,817$), and applied the global predictive residual outlier tests discussed in Section \ref{sec:outlier.test}.  Using only complete observations and setting $n_k=1000$, $m=3$, and $\alpha=0.05$, we found that the normality-based $F$ test in \eqref{eq:f.test} and asymptotic $F$ test in \eqref{eq:f.test.asymp} overwhelmingly agreed upon whether or not there was at least one outlier in a given subset of data (96\% agreement across $K=12803$ subsets).  As in the simulations, the normality-based $F$ test rejects more often than the asymptotic $F$ test: in the 4\% of subsets in which the two tests did not agree, the normality-based $F$ test alone identified 488 additional subsets with at least one outlier, while the asymptotic $F$ test alone identified 23 additional subsets with at least one outlier.


\section{Discussion}

We developed online-updating algorithms and inferences applicable for linear models and estimating equations.
We used the divide and conquer approach to motivate our online-updated estimators for the regression coefficients, and similarly introduced
online-updated estimators for the variances of the regression coefficients.  The variance estimation allows for online-updated inferences.  We note that if one wishes to perform sequential testing, this would require an adjustment of the $\alpha$ level to account for multiple testing.

In the linear model setting, we provided a method for outlier detection using predictive residuals.  Our simulations suggested that the predictive residual tests are more powerful than a test that uses only the current dataset in the stream. In the EE setting, we may similarly consider outlier tests also based on standardized predictive residuals.  For example in generalized linear models, one may consider the sum of squared predictive Pearson or Deviance residuals, computed using the coefficient estimate from the cumulative data (i.e., $\tilde{\bbeta}_{k-1}$ or $\hat{\bbeta}_{k-1}$). It remains an open question in both settings, however, regarding how to handle such outliers when they are detected.  This is an area of future research.

In the estimating equation setting, we also proposed a new online-updated estimator of the regression coefficients that borrows information from previous datasets in the data stream.  The simulations indicated that in finite samples, the proposed CUEE estimator is less biased than the AEE/CEE estimator of \cite{Lin:Xi:aggr:2011}.  However, both estimators were shown to be asymptotically consistent.

The methods in this paper were designed for small to moderate covariate dimensionality $p$, but large $N$.  The use of penalization in the large $p$ setting is an interesting consideration, and has been explored in the divide-and-conquer context in \cite{ChenXie:2014} with popular sparsity inducing penalty functions.  In our online-updating framework, inference for the penalized parameters would be challenging, however, as the computation of variance estimates for these parameter estimates is quite complicated and is also an area of future work.

The proposed online-updating methods are particularly useful for data that is obtained sequentially and without access to historical data. Notably, under the normal linear regression model, the proposed scheme does not lead to any information loss for inferences involving $\bbeta$, as when the design matrix is of full rank, \eqref{betahatk} and \eqref{msenkk} are sufficient and complete statistics for $\bbeta$ and $\sigma^2.$  However, under the estimating equation setting, some information will be lost. Precisely how much information needs to be retained at each subset for specific types of inferences is an open question, and an area devoted for future research.


\bibliographystyle{asa}
\bibliography{bigdata}

\newpage

\setcounter{page}{1}
\def\thepage{S\arabic{page}}

\doublespace

\begin{center}
{\bf
{\Large Supplementary Material: Online Updating of Statistical Inference in the Big Data Setting}}
\end{center}

\section*{A: Bayesian Insight into Online Updating}
\setcounter{equation}{0}
\renewcommand{\theequation}{A.\arabic{equation}}

A Bayesian perspective provides some insight into how we may construct our online-updating estimators. Under a Bayesian framework, using the previous $k-1$ subsets of data to construct a prior distribution for the current data in subset $k$, we immediate identify the appropriate online updating formulae for estimating the regression coefficients and the error variance.  Conveniently, these formulae require storage of only a few low-dimensional quantities computed only within the current subset; storage of these quantities is not required across all subsets.

We first assume a joint conjugate prior for $(\bbeta,\sigma^2)$ as follows:
 \begin{equation}
 \pi( \bbeta,\sigma^2|\bmu_0, \V_0, \nu_0,\tau_0) = \pi(\bbeta|\sigma^2,\bmu_0, \V_0) \pi(\sigma^2|\nu_0,\tau_0), \label{conjprior}
 \end{equation}
 where  $\bmu_0$ is a prespecified $p$-dimensional vector, $\V_0$ is a $p\times p$ positive definite precision matrix,
  $\nu_0 > 0$, $\tau_0 > 0$, and
 \begin{eqnarray*}
   \pi(\bbeta|\sigma^2,\bmu_0, \V_0)&=&\frac{ |\V_0|^{1/2} }{ (2 \pi \sigma^2)^{p/2}}
   \exp\Big\{ - \frac{1}{2 \sigma^2} (\bbeta-\bmu_0)'\V_0 (\bbeta-\bmu_0) \Big\}, \label{normal}\\
 \pi( \sigma^2|\nu_0, \tau_0) &\propto& (\sigma^2)^{-( \nu_0/2 +1) }  \exp \Big\{- \frac{\tau_0}{2 \sigma^2} \Big\}. \label{IG}
 \end{eqnarray*}

 When the data $D_1=\{(\y_1, \X_1)\}$ is available, the likelihood is given by
 \begin{equation*}
 L(\bbeta,\sigma^2|D_1) \propto  \frac{1}{(\sigma^2)^{n_1/2}} \exp \Big\{ -\frac{1}{2\sigma^2} ( \y_1-\X_1\bbeta)'(\y_1-\X_1\bbeta) \Big\}.
 \end{equation*}
 After some algebra, we can show that
 the posterior distribution of $(\bbeta,\sigma^2)$ is then given by
 \begin{equation*}
 \pi( \bbeta,\sigma^2|D_1,\bmu_0, \V_0, \nu_0,\tau_0)= \pi(\bbeta|\sigma^2,\bmu_1,\V_1)
 \pi(\sigma^2|\nu_1,\tau_1),   \label{post1}
 \end{equation*}
 where $\bmu_1 = (\X_1'\X_1+\V_0)^{-1} ( \X_1'\X_1\hat{\bbeta}_{n_1, 1} + \V_0 \bmu_0),$
 $\V_1 =   \X_1'\X_1+\V_0,$
 $\nu_1=   n_1+\nu_0, $ and
$\tau_1 = \tau_0 + (\y_1-\X_1\hat{\bbeta}_{n_1, 1})'(\y_1-\X_1\hat{\bbeta}_{n_1, 1})
       + \bmu_0'\V_0\bmu_0 + \hat{\bbeta}_{n_1, 1}'\X_1'\X_1\hat{\bbeta}_{n_1, 1} - \bmu_1'\V_1\bmu_1;$ see, for example, Section 8.6 of \cite{DeGroot2012}.
Using mathematical induction, we can show that
given the data $D_k=\{(\y_\ell, \X_\ell),\; \ell=1,2,\dots,k\}$, the posterior distribution
of $(\bbeta,\sigma^2)$ is $
  \pi( \bbeta,\sigma^2|\bmu_k, \V_k, \nu_k,\tau_k),
 $
 which has the same form as in \eqref{conjprior} with $(\bmu_0, \V_0, \nu_0,\tau_0)$ updated by
  $(\bmu_k, \V_k, \nu_k,\tau_k)$,
 where
\begin{equation}
    \begin{array}{rl}
  \bmu_k = & (\X_k'\X_k+\V_{k-1})^{-1} ( \X_k'\X_k\hat{\bbeta}_{n_k, k} + \V_{k-1} \bmu_{k-1}), \\
  \V_k = &  \X_k'\X_k+\V_{k-1}, \\
  \nu_k= &  n_k+\nu_{k-1}, \\
\tau_k=& \tau_{k-1} + (\y_k-\X_k\hat{\bbeta}_{n_k, k})'(\y_k-\X_k\hat{\bbeta}_{n_k, k})\\
       &+ \bmu_{k-1}'\V_{k-1}\bmu_{k-1} + \hat{\bbeta}_{n_k, 1}'\X_k'\X_k\hat{\bbeta}_{n_k, k} - \bmu_k'\V_k\bmu_k,\\
  \end{array} \label{Bupdatek}
 \end{equation}
for $k=1,2,\dots$. The data stream structure fits the Bayesian paradigm perfectly and the Bayesian online updating sheds light on the online updating of LS estimators.
Let $\hat{\bbeta}_k$ and MSE$_k$ denote the LS estimate of $\bbeta$
and the corresponding MSE based on
the cumulative data $D_k=\{(\y_\ell, \X_\ell),\; \ell=1,2,\dots,k\}$.
As a special case of Bayesian online update, we can derive the online updates of  $\hat{\bbeta}_k$
and MSE$_k$.  Specifically, we take $\hat{\bbeta}_1=\hat{\bbeta}_{n_1,1}$ and use the updating formula for $\bmu_k$ in \eqref{Bupdatek}.
That is, taking $\bmu_0=\bzero$ and $\V_0=\bzero_p$ in \eqref{Bupdatek}, we obtain
\begin{equation}
\hat{\bbeta}_k=(\X_k'\X_k+\V_{k-1})^{-1} (\X_k'\X_k \hat{\bbeta}_{n_k k} + \V_{k-1} \hat{\bbeta}_{k-1}),
 \label{updatedLSbetak}
\end{equation}
where $\hat{\bbeta}_0=\bzero$, $\hat{\bbeta}_{n_k k}$ is defined by
\eqref{betahatk} or \eqref{eq:ginvbhat} and $\V_k=\sum^k_{\ell=1} \X'_\ell \X_\ell$ for  $k=1,2,\dots$.

Similarly, taking $\nu_0=n_0=0,$ $\tau_0=\mbox{SSE}_0=0$, and using the updating formula for $\tau_k$ in \eqref{Bupdatek}, we have
\begin{eqnarray}
\mbox{SSE}_{k}
 & = & \mbox{SSE}_{k-1} + \mbox{SSE}_{n_k,k}
 + \hat{\bbeta}_{k-1}'\V_{k-1}\hat{\bbeta}_{k-1}+ \hat{\bbeta}_{n_k, k}'\X_k'\X_k\hat{\bbeta}_{n_k, k} - \hat{\bbeta}_{k}'\V_k\hat{\bbeta}_{k}  \label{MSEupdatek.variant}
\end{eqnarray}
where SSE$_{n_k,k}$ is the residual sum of squares from the $k^{th}$ dataset, with corresponding residual mean square MSE$_{n_k,k}=$SSE$_{n_k,k}/(n_k-p).$
The MSE based on the data $D_k$ is then $ \mbox{MSE}_k=\mbox{SSE}_k/(N_k-p)$
where $N_k=\sum^k_{\ell=1} n_\ell$ $(=n_k+N_{k-1})$ for $k=1,2,\ldots$.  Note that for $k=K$, equations \eqref{updatedLSbetak} and \eqref{MSEupdatek.variant} are identical to those in \eqref{Ksubsetbeta} and \eqref{MSEform}, respectively.

\section*{B: Online Updating Statistics in Linear Models}
\setcounter{equation}{0}
\renewcommand{\theequation}{B.\arabic{equation}}

Below we provide online-updated $t$-tests for the regression parameter estimates, the online-updated ANOVA table, and online-updated general linear hypothesis $F$-tests.  Please refer to Section \ref{sec:lsest} of the main text for the relevant notation.

\medskip

\noindent
{\bf Online Updating for Parameter Estimate t-tests in Linear Models}.
If our interest is only in performing $t$-tests for the regression coefficients, we only need to save the current values $(\V_{k},\hat{\bbeta}_{k}, N_{k}, \mbox{MSE}_{k})$ to proceed.  Recall that $\mbox{var}(\hat{\bbeta}) = \sigma^2(\X'\X)^{-1}$ and $\widehat{\mbox{var}}(\hat{\bbeta}) = \mbox{MSE}(\X'\X)^{-1}.$  At the $k^{th}$ update, $\widehat{\mbox{var}}(\hat{\bbeta}_k) = \mbox{MSE}_k \V_k^{-1}.$
Thus, to test $H_0: \beta_j=0$ at the $k^{th}$ update ($j=1,\ldots,p$), we may use $ t^\ast_{k,j} = \hat{\beta}_{k,j}/ se(\hat{\beta}_{k,j}), $
where the standard error $se(\hat{\beta}_{k,j})$ is the square root of the $j^{th}$ diagonal element of $\widehat{\mbox{var}}(\hat{\bbeta}_k).$
The corresponding p-value is $P(|t_{N_k-p}| \geq |t^\ast_{k,j}|)$.

\medskip
\noindent {\bf Online Updating for ANOVA Table in Linear Models}.
Observe that SSE is given by \eqref{MSEform},
$$
\mbox{SST} = \y'\y - N\bar{y}^2=\sum_{k=1}^K\y_k'\y_k - N^{-1}(\sum_{k=1}^K\y_k'\1_{n_k})^2,
$$
where $\1_{n_k}$ is an $n_k$ length vector of ones, and SSR = SST-SSE.  If we wish to construct an online-updated ANOVA table, we must save two additional easily computable, low dimensional quantities: $S_{yy,k}= \sum_{\ell=1}^k\y_\ell'\y_\ell$ and $S_{y,k} = \sum_{\ell=1}^k\y_\ell'\1_{n_\ell} =\sum_{\ell=1}^k\sum_{i=1}^{n_\ell}y_{\ell i}.$

The online-updated ANOVA table at the $k^{th}$ update for the cumulative data $D_k$ is constructed as in Table \ref{tab:anova}.
Note that SSE$_k$ is computed as in \eqref{MSEupdatek.variant}.  The table may be completed upon determination of an updating formula SST$_k.$
Towards this end, write $S_{yy,k}= \y_k'\y_k + S_{yy,k-1}$ and $S_{y,k} = \y_k'1_{n_k} + S_{y,k-1},$ for $k=1,\ldots,K$ and $S_{yy,0}=S_{y,0}=0,$ so that $\mbox{SST}_k = S_{yy,k} - N_k^{-1}S_{y,k}^2$

\begin{table}[t]
{\singlespace
\caption{Online-updated ANOVA table}\label{tab:anova}
\begin{tabular}{lccccc} \hline
\multicolumn{2}{l}{ANOVA$_{N_k}$ Table} &      &     &    &     \\
Source  &  df     &  SS  & MS  & F  & P-value \\ \hline
Regression & $p-1$   &  SSR$_k$ & MSR$_k=\frac{\mbox{SSR}_k}{p-1}$ & $F^\ast=\frac{\mbox{MSR}_k}{\mbox{MSE}_k}$& $P(F_{p-1,N_k-p} \geq F^\ast)$ \\
Error      & $N_k-p$ &  SSE$_k$ & MSE$_k=\frac{\mbox{SSE}_k}{N_k-p}$ &    &      \\
C Total    & $N_k-1$ & SST$_k$ &    &      &   \\ \hline
\end{tabular}}
\end{table}

Online updated testing of General Linear Hypotheses ($H_0: \C\bbeta=\bzero$) are also possible:
if $\C$ ($q\times p$) is a full rank ($q\leq p$) contrast matrix, under $H_0,$
$$
F_k=(\frac{\hat\bbeta_k'\C'(\C\V_k^{-1}\C')^{-1}\C\hat\bbeta_k}{q})/(\frac{\mbox{SSE}_k}{N_k-p})\sim F_{q,N_k-p}.
$$
Similarly, we may also obtain online updated coefficients of multiple determination, $R^2_k = \mbox{SSR}_k/\mbox{SST}_k.$

To summarize, we need only save $(\V_{k-1},\hat{\bbeta}_{k-1}, N_{k-1}, \mbox{MSE}_{k-1},S_{yy,k-1},S_{y,k-1})$ from the previous accumulation point $k-1$ to perform online-updated $t$-tests for $H_0:\beta_j=0,$ $j=1,\ldots,p$ and online-updated $F$-tests for the current accumulation point $k$; we do not need to retain $(\V_{\ell},\hat{\bbeta}_{\ell}, N_{\ell}, \mbox{MSE}_{\ell},S_{yy,\ell},S_{y,\ell})$ for $\ell=1,\ldots,k-2.$

\section*{C: Proof of Proposition \ref{prop:chisq}}\label{sup:asympF}
\setcounter{equation}{0}
\renewcommand{\theequation}{C.\arabic{equation}}
We first show that $\mbox{MSE}_{k-1} \xrightarrow[]{p}\sigma^2$. Since $\mbox{SSE}_{k-1}=\bvarepsilon_{k-1}'(\I_{N_{k-1}}-\mcX_{k-1}(\mcX_{k-1}'\mcX_{k-1})^{-1}\mcX_{k-1}')\bvarepsilon_{k-1}$, we have
\begin{align*}
\plim_{N_{k-1} \rightarrow \infty}\mbox{MSE}_{k-1}&=\plim_{N_{k-1} \rightarrow \infty}\frac{\mbox{SSE}_{k-1}}{N_{k-1}-p}\\
&=\plim_{N_{k-1} \rightarrow \infty}\frac{\bvarepsilon_{k-1}'\bvarepsilon_{k-1}}{N_{k-1}} - \plim_{N_{k-1} \rightarrow \infty} \frac{\bvarepsilon_{k-1}'\mcX_{k-1}(\mcX_{k-1}'\mcX_{k-1})^{-1}\mcX_{k-1}'\bvarepsilon_{k-1}}{N_{k-1}}\\
&=\sigma^2-\plim_{N_{k-1} \rightarrow \infty}{\frac{\bvarepsilon_{k-1}'\mcX_{k-1}}{N_{k-1}}}\plim_{N_{k-1} \rightarrow \infty}{(\frac{\mcX_{k-1}'\mcX_{k-1}}{N_{k-1}})}^{-1}\plim_{N_{k-1} \rightarrow \infty}{\frac{\mcX_{k-1}'\bvarepsilon_{k-1}}{N_{k-1}}}
\end{align*}
Let $\mathcal{X}_j$ denote the column vector of $\mcX_{k-1}$, for $j=1, \ldots, p$. Since $E(\epsilon_i)=0$, $\forall i$ and all the elements of $\mcX_{k-1}$ are bounded by $C$, by Chebyshev's Inequality we have for any $\ell$ and column vector $\mathcal{X}_j$,
$$
P(|\frac{\bvarepsilon_{k-1}'\mathcal{X}_j}{N_{k-1}}|\geq \ell)\le \frac{Var(\bvarepsilon_{k-1}'\mathcal{X}_j)}{\ell^2N^2_{k-1}}\le \frac{C^2\sigma^2}{\ell^2 N_{k-1}},
$$
and thus $\displaystyle \plim_{N_{k-1} \rightarrow \infty}{\frac{\bvarepsilon_{k-1}'\mcX_{k-1}}{N_{k-1}}}=0$ and
$$
\plim_{N_{k-1} \rightarrow \infty}\mbox{MSE}_{k-1}=\sigma^2 - 0 \cdot \Q^{-1} \cdot 0=\sigma^2.
$$
Next we show $\frac{\sum_{i=1}^m \frac{1}{n_{k_i}}({\bf 1}_{k_i} '\check{\e}^*_{k_i})^2}{\sigma^2} \xrightarrow[]{d} \chi^2_{m}$. First, recall that
\begin{align*}
\check{\e}_{k}&=\y_k-\check{\y}_k\\
&=\X_k\bbeta+\bepsilon_k-\X_k\hat{\bbeta}_{k-1}\\
&=\X_k\bbeta+\bepsilon_k-\X_k(\mcX_{k-1}'\mcX_{k-1})^{-1}\mcX_{k-1}'\mcy_{k-1}\\
&=\bepsilon_k-\X_k(\mcX_{k-1}'\mcX_{k-1})^{-1}\mcX_{k-1}'\bvarepsilon_{k-1}.
\end{align*}
Consequently, $\mbox{var}(\check{\e}_{k})=(\I_{n_k}+\X_k(\mcX_{k-1}'\mcX_{k-1})^{-1}\X_{k}')\sigma^2\triangleq \bGamma'\bGamma\sigma^2,$ where $\bGamma$ is an $n_k \times n_k$ invertible matrix. Let $\check{\e}^*_k=(\bGamma')^{-1}\check{\e}_k$ with $\mbox{var}(\check{\e}^*_k) = \sigma^2\I_{n_k}$. Therefore, each component of $\check{\e}^*_k$ is independent and identically distributed.\\
By the Central Limit Theorem and condition (4), we have for all $i=1, \ldots, m$,
\begin{equation*}
\frac{\frac{1}{n_{k_i}}({\bf 1}_{k_i} '\check{\e}^*_{k_i})^2}{\sigma^2} \xrightarrow[]{d} \chi^2_{1}, \text{ \qquad as $n_k \rightarrow \infty$.}
\end{equation*}
Since each subgroup is also independent,
\begin{equation*}
\frac{\sum_{i=1}^m \frac{1}{n_{k_i}}({\bf 1}_{k_i} '\check{\e}^*_{k_i})^2}{\sigma^2} \xrightarrow[]{d} \chi^2_{m}, \text{ \qquad as $n_k \rightarrow \infty$.}
\end{equation*}
By Slutsky's theorem,
 \begin{equation*}
\frac{\sum_{i=1}^m \frac{1}{n_{k_i}}({\bf 1}_{k_i} '\check{\e}^*_{k_i})^2}{\mbox{MSE}_{k-1}} \xrightarrow[]{d} \chi^2_{m}, \text{ \qquad as $n_k, N_{k-1} \rightarrow \infty$.}
 \end{equation*}
  \hfill$\blacksquare$

\section*{D: Computation of $\bGamma$ for Asymptotic F test}\label{sup:proof}
\setcounter{equation}{0}
\renewcommand{\theequation}{D.\arabic{equation}}

Recall that $\mbox{var}(\check{\e}_{k})=(\I_{n_k}+\X_k(\mcX_{k-1}'\mcX_{k-1})^{-1}\X_{k}')\sigma^2\triangleq \bGamma\bGamma'\sigma^2,$ where $\bGamma$ is an $n_k \times n_k$ invertible matrix. For large $n_k$, it may be challenging to compute the Cholesky decomposition $\mbox{var}(\check{\e}_{k})$. One possible solution that avoids the large $n_k$ issue is given as follows.

First, we can easily obtain the Cholesky decomposition of $(\mcX_{k-1}'\mcX_{k-1})^{-1}=\V_{k-1}^{-1}\triangleq \bfP'\bfP$ since it is a $p \times p$ matrix. Thus, we have
\begin{equation*}
\mbox{var}(\check{\e}_{k})=(\I_{n_k}+\X_k\bfP'\bfP\X_{k}')^{-1}\sigma^2=(\I_{n_k}+\tilde{\X}_k\tilde{\X}_{k}')^{-1}\sigma^2,
\end{equation*}
where $\tilde{\X}_k=\X_k\bfP'$ is an $n_k \times p$ matrix.

Next, we compute the singular value decomposition on $\tilde{\X}_k$, i.e., $\tilde{\X}_k=\U\D\V'$ where $\U$ is an $n_k \times n_k $ unitary matrix, $\D$ is an $n_k  \times n_k $ diagonal matrix, and $\V$ is a $n_k  \times p$ unitary matrix. Therefore,
\begin{equation*}
\mbox{var}(\check{\e}_{k})=(\I_{n_k}+ \U\D\D'\U')^{-1}\sigma^2=\U(\I_{n_k}+\D\D')^{-1}\U'\sigma^2
\end{equation*}
Since $(\I_{n_k}+\D\D')^{-1}$ is a diagonal matrix, we can find the matrix $\Q$ such that $(\I_{n_k}+\D\D')^{-1}\triangleq \Q'\Q$ by straightforward calculation. One possible choice of $\bGamma$ is $\U\Q'$.

\section*{E: Proof of Theorem \ref{thm:cuee_consistent}}\label{sup:proof}
\setcounter{equation}{0}
\renewcommand{\theequation}{E.\arabic{equation}}

We use the same definition and two facts provided by \cite{Lin:Xi:aggr:2011}, given below for completeness.

\begin{define}
Let $\A$ be a $d \times d$ positive definite matrix. The
norm of $\A$ is defined as $\left\|\A\right\|= \mbox{sup}_{\bv\in\Real^d,\bv\neq \bzero} \frac{\left\|\A\bv\right\|}{\bv}.$
\end{define}

Using the definition of the above matrix norm, one may verify the following two facts.

\noindent{\bf Fact A.1.} Suppose that $\A$ is a $d\times d$ positive definite matrix.
Let $\lambda$ be the smallest eigenvalue of $\A$, then we have $\bv'\A\bv \geq \lambda\bv'\bv = \lambda\left\|\bv\right\|^2$ for any vector $\bv\in\Real^d.$ On the contrary, if there exists a constant $C > 0$ such that $\bv'\A\bv \geq C\left\|\bv\right\|^2$ for
any vector $\bv\in\Real^d,$ then $C\leq \lambda.$

\noindent{\bf Fact A.2.} Let $\A$ be a $d\times d$ positive definite matrix and $\lambda$ is
the smallest eigenvalue of $\A.$ If $\lambda \geq  c > 0$ for some constant
$c,$ one has $\left\| \A^{-1} \right\| \leq c^{-1}.$

In order to prove Theorem \ref{thm:cuee_consistent}, we need the following Lemma.

\begin{lem}\label{lem:c6}
Under (C4') and (C6),  $\check\bbeta_{n,k}$  satisfies the following condition: for any $\eta >0$, $\displaystyle n^{-2\alpha+1}P(n^{\alpha}\|\check\bbeta_{n,k}-\bbeta_{0}\| > \eta)=O(1)$.
\end{lem}

\noindent\textbf{Proof of Lemma \ref{lem:c6} (By induction)}\\
First notice that (C6) is equivalent to writing, for any $\eta >0$, $\displaystyle n^{-2\alpha+1}P(n^{\alpha}\|\hat\bbeta_{n,k}-\bbeta_{0}\| > \eta)=O(1)$.\\
Take $k=1$,  $\check\bbeta_{n,1}=\hat\bbeta_{n, 1}$ and thus $\displaystyle n^{-2\alpha+1}P(n^{\alpha}\|\check\bbeta_{n,1}-\bbeta_{0}\| > \eta)=O(1)$.

\noindent Assume the condition holds for accumulation point $k-1$:  $n^{-2\alpha+1}P(n^{\alpha}\|\check\bbeta_{n,k-1}-\bbeta_{0}\| > \eta)=O(1).$
Write
\begin{align*}
\check\bbeta_{n,k-1}&= (\tilde{\A}_{k-2} + \A_{n,k-1})^{-1}
(\sum_{\ell=1}^{k-2}\tilde\A_{n,\ell}\check\bbeta_{n,\ell} + \A_{n,k-1} \hat\bbeta_{n,k-1})\\
\shortintertext{so that, rearranging terms, we have}
\sum_{\ell=1}^{k-2}\tilde\A_{n,\ell}\check\bbeta_{n,\ell}&=
(\tilde{\A}_{k-2} + \A_{n,k-1})\check\bbeta_{n,k-1} - \A_{n,k-1} \hat\bbeta_{n,k-1}.\\
\shortintertext{Using the previous relation, we may write $\check\bbeta_{n,k}$ as}
\check\bbeta_{n,k}&=(\tilde{\A}_{k-1} + \A_{n,k})^{-1}
(\tilde{\A}_{k-2}\check\bbeta_{n,k-1} + \tilde\A_{n,{k-1}}\check\bbeta_{n,k-1} + \\
&\qquad\A_{n,k} \hat\bbeta_{n,k} + \A_{n,k-1} (\check\bbeta_{n,k-1}-\hat\bbeta_{n, {k-1}}))\\
&=(\tilde{\A}_{k-1} + \A_{n,k})^{-1}
(\tilde{\A}_{k-1}\check\bbeta_{n,k-1} + \A_{n,k} \hat\bbeta_{n,k} + \A_{n,k-1} (\check\bbeta_{n,k-1}-\hat\bbeta_{n, {k-1}})).\\
\shortintertext{Therefore,}
\check\bbeta_{n,k} - \bbeta_{0}&=(\tilde{\A}_{k-1} + \A_{n,k})^{-1}
(\tilde{\A}_{k-1}(\check\bbeta_{n,k-1} - \bbeta_{0}) + \A_{n,k} (\hat\bbeta_{n,k} - \bbeta_{0}) + \\
 & \qquad\A_{n,k-1} (\check\bbeta_{n,k-1} - \bbeta_{0} + \bbeta_{0} -\hat\bbeta_{n, {k-1}}))\\
\shortintertext{and}
 \|\check\bbeta_{n,k} - \bbeta_{0}\|
 & \le \|(\tilde{\A}_{k-1} + \A_{n,k})^{-1}\tilde{\A}_{k-1}\|\|\check\bbeta_{n,k-1}-\bbeta_{0}\| + \\
 &\quad \|(\tilde{\A}_{k-1} + \A_{n,k})^{-1}\A_{n,k}\|\|\hat\bbeta_{n, k}-\bbeta_{0}\| + \\
 &\quad \|(\tilde{\A}_{k-1} + \A_{n,k})^{-1}\A_{n,k-1}\|\|\check\bbeta_{n,k-1}-\bbeta_{0}\| + \\
 &\quad \|(\tilde{\A}_{k-1} + \A_{n,k})^{-1}\A_{n,k-1}\|\|\hat\bbeta_{n, {k-1}}-\bbeta_{0}\|\\
 \shortintertext{Note that $\|(\tilde{\A}_{k-1} + \A_{n,k})^{-1}\tilde{\A}_{k-1}\|\le 1$ and $\|(\tilde{\A}_{k-1} + \A_{n,k})^{-1}\A_{n,k}\|\le 1$.  Under (C4'), $\|(\tilde{\A}_{k-1} + \A_{n,k})^{-1}\A_{n,k-1}\|
 \le \|(\A_{n,k})^{-1}\A_{n,k-1}\|
 \le\frac{\lambda_2}{\lambda_1}
 \le C$, where $C$ is a constant, $\lambda_1>0$ is the smallest eigenvalue of $\bLambda_1$, and $\lambda_2$ is the largest eigenvalue of $\bLambda_2$.  Note that if  $n_k\neq n$ for all $k$, then
 $\|(\tilde{\A}_{k-1} + \A_{n,k})^{-1}\A_{n,k-1}\|
 \le \|(\A_{n,k})^{-1}\A_{n,k-1}\|
 \le\frac{n_{k-1}}{n_k}\frac{\lambda_2}{\lambda_1}
 \le C$, where $n_{k-1}/n_k$ is bounded and $C$ is a constant. Thus,
 }
\|\check\bbeta_{n,k} - \bbeta_{0}\|
 &\le \|\check\bbeta_{n,k-1}-\bbeta_{0}\| + \|\hat\bbeta_{n, k}-\bbeta_{0}\| +\\
 &\quad\|C(\check\bbeta_{n,k-1}-\bbeta_{0})\| + \|C(\hat\bbeta_{n, {k-1}}-\bbeta_{0})\|\\
 \shortintertext{Under (C6) and the induction hypothesis, then for any $\eta >0$,}
 \displaystyle n^{-2\alpha+1}P(\|\check\bbeta_{n,k}-\bbeta_{0}\| > \frac{\eta}{n^{\alpha}})
 &\le\displaystyle n^{-2\alpha+1}P(\|\check\bbeta_{n,k-1}-\bbeta_{0}\| > \frac{\eta}{4 n^{\alpha}})  +\\
 &\displaystyle n^{-2\alpha+1}P(\|\hat\bbeta_{n, k}-\bbeta_{0}\| > \frac{\eta}{4 n^{\alpha}}) + \\
 &\displaystyle n^{-2\alpha+1}\quad P(\|\check\bbeta_{n,k-1}-\bbeta_{0}\| > \frac{\eta}{4 C n^{\alpha}}) +\\
 &\displaystyle n^{-2\alpha+1}P(\|\hat\bbeta_{n, {k-1}}-\bbeta_{0}\| > \frac{\eta}{4 C n^{\alpha}})
  \end{align*}
Since all the four terms on the right hand side are $O(1)$ by assumption,
 $\displaystyle n^{-2\alpha+1}P(\|\check\bbeta_{n,k}-\bbeta_{0}\| > \frac{\eta}{n^{\alpha}})=O(1)$.\hfill$\blacksquare$

 We are now ready to prove Theorem \ref{thm:cuee_consistent}:\\
 \noindent\textbf{Proof of Theorem \ref{thm:cuee_consistent}}\\
 First, suppose that all the random variables are defined on a probability space $(\Omega, \mathcal{F}, \mathbb{P})$. Let
\begin{align*}
\Omega_{n, k, \eta}&=\{\omega|n^{\alpha}\|\check\bbeta_{n,k}-\bbeta_{0}\|\le \eta\},\\
\Omega_{N, \eta}&=\{\omega|N^{\alpha}\|\hat\bbeta_{N}-\bbeta_{0}\|\le \eta\},\\
\Gamma_{N, k, \eta}&=\cap_{k=1}^K\Omega_{n, k, \eta}\cap\Omega_{N, \eta}.\\
\shortintertext{From Lemma \ref{lem:c6}, for any $\omega>0$, we have}
\displaystyle P(\Gamma_{N, k, \eta}^c)
& \le P(\Omega_{N, \eta}^c)+ \sum_{k=1}^{K}P(\Omega_{n, k, \eta}^c)\\
& \le n^{2\alpha-1}(O(1)+K \cdot O(1))
\end{align*}
Since K=$O(n^{\gamma})$,  $\gamma < 1- 2\alpha$ and $\frac{1}{4}\le\alpha\le\frac{1}{2}$  by assumption, we have $\displaystyle\lim_{n\to\infty}P(\Gamma_{N, k, \eta}^c)\rightarrow 0.$

Next, we wish to show $\Gamma_{N, k, \eta} \subseteq \{\omega|\sqrt{N}\|\hat\bbeta_N-\tilde\bbeta_K\|\le \delta\}.$  Consider the Taylor expansion of $-M_{n,k}(\hat\bbeta_N)$ at intermediary estimator $\check\bbeta_{n,k}:$
\begin{align*}
-M_{n,k}(\hat\bbeta_N) &= -M_{n,k}(\check\bbeta_{n,k}) + [\A_{n,k}(\check\bbeta_{n,k})](\hat\bbeta_N-\check\bbeta_{n,k}) + \check\bfr_{n,k},\\
\shortintertext{where $\check\bfr_{n,k}$ is the remainder term with $j^{th}$ element    $\frac{1}{2}(\hat\bbeta_N-\check\bbeta_{n,k})'\sum_{i=1}^n\frac{-\partial^2\psi_j(\z_{ki}, \bbeta_k^*)}{\partial\bbeta\partial\bbeta'}(\hat\bbeta_N-\check\bbeta_{n,k})$ for some $\bbeta_k^*$  between $\hat\bbeta_N$ and $\check\bbeta_{n,k}.$ }
\shortintertext{Summing over $k$,}
0 &= -\sum_{k=1}^K{M_{n,k}(\hat\bbeta_N)}=-\sum_{k=1}^K M_{n,k}(\check\bbeta_{n,k})+\sum_{k=1}^K\A_{n,k}(\check\bbeta_{n,k})(\hat\bbeta_N-\check\bbeta_{n,k}) + \sum_{k=1}^K\check\bfr_{n,k}.\\
\shortintertext{Rearranging terms and recalling that $\A_{n,k}(\check\bbeta_{n,k})=\tilde\A_{n,k},$ we find }
-\hat\bbeta_N &+ (\sum_{k=1}^K\tilde\A_{n,k})^{-1}(\sum_{k=1}^K\tilde\A_{n,k}\check\bbeta_{n,k} + \sum_{k=1}^KM_{n,k}(\check\bbeta_{n,k})) = (\sum_{k=1}^K\tilde\A_{n,k})^{-1}\sum_{k=1}^K\check\bfr_{n,k}.\\
\shortintertext{Using the definition of the CUEE estimator $\tilde\bbeta_{K}$, the above relation reduces to}
\tilde\bbeta_{K}-\hat\bbeta_N&=(\sum_{k=1}^K\tilde\A_{n,k})^{-1}\sum_{k=1}^K\check\bfr_{n,k}\\
\shortintertext{and}
\|\tilde\bbeta_{K}-\hat\bbeta_{N}\| & \le
\left\|\left(\frac{1}{nK}\sum_{k=1}^K\tilde{\A}_{n,k}\right)^{-1}\right\| \left\|\frac{1}{nK}\sum_{k=1}^K\check\bfr_{n,k}\right\|.
\end{align*}
For the first term, according to (C4'), $\left\|\left(\frac{1}{nK}\sum_{k=1}^K\tilde{\A}_{n,k}\right)^{-1}\right\| \le \lambda_1^{-1}$ since ${\A}_{n,k}(\bbeta)$ is a continuous function of $\bbeta$ (according to (C1)) and $\check\bbeta_{n,k}$ is in the neighborhood of $\bbeta_0$ for small enough $\eta$. For the second term, we introduce set $B_{\eta}(\bbeta_0)=\{\bbeta|\|\bbeta-\bbeta_0\|\le\eta\}$.  For all $\omega\in\Gamma_{N,k,n},$ we have  $\bbeta^*_k \in B_{\eta}(\bbeta_0)$ since $B_{\eta}(\bbeta_0)$ is a convex set and $\hat\bbeta_N, \check\bbeta_{n,k}\in B_{\eta}(\bbeta_0).$
According to (C5), for small enough $\eta$, $B_{\eta}(\bbeta_0)$ satisfies (C5) and thus $\bbeta^*_k$ satisfies (C5).  Hence we have
$\|\check\bfr_{n,k}\| \le C_2pn\|\hat\bbeta_N-\check\bbeta_{n,k}\|^2$ for all $\omega\in\Gamma_{N,K,\eta}$ when $\eta$ is small enough.
Additionally,
\begin{align*}
\|\check\bfr_{n,k}\|& \le C_2pn\|\hat\bbeta_N-\check\bbeta_{n,k}\|^2 \le C_2pn(\|\hat\bbeta_N-\bbeta_0\|^2 + \|\check\bbeta_{n,k}-\bbeta_0\|^2)\\
&\le C_2pn(\frac{\eta^2}{n^{2\alpha}}+\frac{\eta^2}{N^{2\alpha}})\\
&\le 2C_2pn^{1-2\alpha}{\eta}^2.
\end{align*}
Consequently,
$$
\|\tilde\bbeta_{K}-\hat\bbeta_{N}\|  \le \frac{1}{\lambda_1}\frac{K}{nK}2c_2pn^{1-2\alpha}{\eta}^2 \le C\frac{\eta^2}{n^{2\alpha}},
$$
where $C=\frac{2C_2p}{\lambda_1}.$

Therefore, for any $\delta>0,$ there exists $\eta_\delta >0$ such that $C\eta_\delta^2 < \delta$.  Then for any $\omega \in \Gamma_{N, k, \eta_\delta}$ and $K=O(n^\gamma)$, where $\gamma<min\{1-2\alpha, 4\alpha-1\}$,  we have $\sqrt{N}\|\tilde\bbeta_{K}-\hat\bbeta_{N}\|\le O(n^{\frac{1+\gamma-4\alpha}{2}})\delta.$  Therefore, when $n$ is large enough, $\Gamma_{N, k, \eta} \subseteq \{\omega\in\Omega|\sqrt{N}\|\hat\bbeta_N-\tilde\bbeta_K\|\le \delta\}$
 and thus $P(\sqrt{N}\|\tilde\bbeta_{K}-\hat\bbeta_{N}\| > \delta)\le P(\Gamma_{N, k, \eta}^c)\rightarrow 0$ as $n\rightarrow\infty.$
 \hfill$\blacksquare$
 
\section*{F: Proof of Proposition \ref{prop:glm}}\label{sup:proof_irls}
\setcounter{equation}{0}
\renewcommand{\theequation}{F.\arabic{equation}}

Suppose $\X_k$ does not have full column rank for some accumulation point $k$.  For ease of exposition, write $\bar\W_k = \mbox{Diag}\bigg(S_{ki}^2W_{ki}\bigg).$  Note that for generalized linear models with $y_{ki}$ from an exponential family, $W_{ki}=1/\mbox{v}(\mu_{ki})$ where $\mbox{v}(\mu_{ki})$ is the variance function. The IRLS approach is then implemented as follows.  For $t=1,2,\ldots$,
\begin{align*}
{\bar\W_k}^{(t)}&=\mbox{Diag}\bigg((S_{ki}^2)^{(t-1)}W_{ki}^{(t-1)}\bigg)\\
\Z^{(t)}_k &=\bfeta_{k}^{(t-1)}+\{\bS_k^{(t-1)}\}^{-1}(\y_k-\bmu_{k}^{(t-1)}) \\
\bbeta^{(t)}&=(\X'_k \bar\W_k^{(t-1)}\X_k)^{-}\X'_k \bar\W_k^{(t-1)}\Z^{(t-1)}_k\\
\bfeta^{(t)}_k&=\X_k\bbeta^{(t)}. 
\end{align*}

As $\X_k$ is not of full rank, $\bbeta^{(t)}$ uses a generalized inverse and is not unique. Since ${\bar\W_k}^{(t)}$ is a diagonal positive definite matrix, there exists an invertible matrix $\bf{V}$ such that ${\bar\W_k}^{(t)}=\bf{V'} \bf{V}$, where $\bf{V}=\sqrt{{\bar\W_k}^{(t)}}$. We thus have
$$
\bfeta^{(t)}_k=\V^{-1} \V \X_k\{(\V \X_k)'(\V \X_k)\}^{-}(\V \X_k)'{\V'}^{-1}{\bar\W_k}^{(t-1)}\Z^{(t-1)}_k.
$$
Therefore, for $t=1$, $\bfeta^{(1)}_k$ is unique no matter what generalized inverse of $\X'_k \bar\W_k^{(0)}\X_k$ we use, given the same initial value $\W_k^{(0)}$. Furthermore,
since ${\bar\W_k}^{(t)}$ and $\Z^{(t)}_k$ depend on $\bbeta^{(t-1)}$ only through $\bfeta^{(t-1)}$, ${\bar\W_k}^{(1)}, \Z^{(1)}_k$ and thus $\bfeta^{(1)}_k$ are also invariant of the choice of generalized inverse. Similarly, we can show that for each iteration, ${\bar\W_k}^{(t)}, \Z^{(t)}_k$ and $\bfeta^{(t)}_k$ are unique no matter what generalized inverse of $\X'_k \bar\W_k^{(t-1)}\X_k$ we use, given the same initial values. 

Now, the only problem left is whether the IRLS algorithm converges. We next show that $\bbeta^{(t)}$ converges under a special generalized inverse of $\X'_k \bar\W_k^{(t-1)}\X_k$.  Let $\X^*_k$ denote a $n_k \times p^*$ full rank column submatrix of $\X_k$. Without loss of generality, we assume the $p^*$ columns of $\X^*_k$ are the first $p^*$ columns of $\X_k$. Assume $\X^*_k$ satisfies (C1-C3) as given in Section \ref{sec:asymptotics}, and the IRLS estimates converge to $\hat{\bbeta}^*_k=({\X^*_k}'\bar\W_k{\X^*_k})^{-1}{\X^*_k}'\bar\W_k\Z_k$, where $\bbeta^*_k$ is the $p^*\times 1$ vector of regression coefficients corresponding to $\X^*_k$. Since $\X^*_k$ is a full column rank submatrix of $\X_k$, there exists a $p^* \times p$ matrix $\bf{P}$ such that $\X_k=\X^*_k \bf{P}$, where the first $p^* \times p^*$ submatrix is an identity matrix. We thus have,
  \begin{align*}
  \bbeta^{(t)}&=(\bfP'{\X^*_k}' \bar\W_k^{(t-1)}\X^*_k \bfP)^{-} \bfP' {\X^*_k}' \bar\W_k^{(t-1)}\Z^{(t-1)}_k\\
  &=
  \begin{pmatrix}
 &{\X^*_k}' \bar\W_k^{(t-1)} {\X^*_k}   &  \bf{0} \\
 & \bf{0}  & \bf{0}   \\
 \end{pmatrix}
 ^{-} \bfP' {\X^*_k}'\bar\W_k^{(t-1)}\Z^{(t-1)}_k\\
 &=
   \begin{pmatrix}
 &{(\X^*_k}' \bar\W_k^{(t-1)} {\X^*_k})^{-1}\\
 & \bf{0}\\
 \end{pmatrix}
 {\X^*_k}'\bar\W_k^{(t-1)}\Z^{(t-1)}_k
  \end{align*}
  Thus, for that special generalized inverse, $\bbeta^{(t)}$ converges to $\big(\hat{\bbeta}^*_k \quad \bf{0}\big)'$. By the uniqueness property given above, $\bbeta^{(t)}$ converges no matter what generalized inverse we choose.

Upon convergence, $\bbeta^{(t)}=\hat\bbeta_{n_k,k}=(\X'_k \bar\W_k\X_k)^{-}\X_k' \bar\W_k \Z_k$ and $\A_{n_k,k}=\X'_k \bar\W_k\X_k$.  As in the normal linear model case,
$\A_{n_k,k} \hat\bbeta_{n_k,k}$ is invariant to the choice of $\A_{n_k,k}^-$, as it is always $\X_k' \W_k \Z_k$.
Therefore, the combined estimator $\hat\bbeta_{NK}$ is invariant to
the choice of generalized inverse $\A_{n_k,k}^-$ of $\A_{n_k,k}$.
Similar arguments can be used for the online estimator $\tilde\bbeta_K$.
 \hfill$\blacksquare$

\end{document}